\documentclass{aa} 
\usepackage{graphicx}

\begin{document}


\title{Chemical composition of Galactic OB stars}
\subtitle{I. CNO abundances in O9 stars\thanks{The INT is operated on the island of La Palma by the RGO in the Spanish Obervatorio de El Roque de los Muchachos of the Instituto de Astrof\'\i sica de Canarias.}}

\author{M.R. Villamariz\inst{1}, A. Herrero\inst{1,2}, S.R. Becker\inst{3} and
K. Butler\inst{3}}

\institute{Instituto de Astrof\'\i sica de Canarias, E-38200 La Laguna, 
Tenerife, Spain
\and
Departamento de Astrof\'{\i}sica, Universidad de La Laguna,
Avda. Astrof\'{\i}sico Francisco S\'anchez, s/n, E-38071 La Laguna, Spain
\and
Universit\"ats-Sternwarte M\"unchen, Scheinerstr. 1, D-81679 M\"unchen, 
Germany
}

\offprints{M.R. Villamariz}

\date{Submitted/Accepted}
\titlerunning{CNO abundances in O9 stars}
  \authorrunning{M.R. Villamariz et al.}

\abstract{We present NLTE abundances of CNO for a sample of four O9
stars in the Galaxy, together with new determinations of their stellar
parameters, $T_{\rm eff}$, $\log g$, $\epsilon$(He) and
microturbulence. These new analyses take into account the effect of
{\it line--blocking} in the spectral synthesis with our classical
NLTE, plane--parallel and hydrostatic model atmospheres.\\ The sample
includes three O9 He normal stars: two dwarfs, HD\,214680 and
HD\,34078, and one supergiant, HD\,209975, and one fast rotating giant
with a preliminary high He overabundance, HD\,191423 with
$\epsilon$(He)=0.20.\\ We find first that the consideration of
microturbulence in the spectral synthesis for the fast rotator leads
to a considerably lower He abundance, $\epsilon$(He)=0.12.\\ The
CNO abundances of the three He normal stars are in good agreement with
the values in the literature for Galactic B dwarfs with no evidence of
mixing, and show that they all have the same chemical composition. We 
also discuss however the possible CNO contamination of the supergiant
HD\,209975. For the fast rotator we find that the abundances show the 
trend of the CNO
contamination: a N overabundance together with C and O depletion.
The N/C and N/O ratios of our stars as a function of their projected
rotational velocities are consistent with the predictions
of the recent evolutionary models of Meynet \& Maeder
\cite{Mey&Mae00}.  
}
\maketitle

\section{Introduction}

Massive blue stars strongly influence the chemical evolution of
galaxies by their interaction with the Interstellar Medium, returning
to it the material from their atmospheres and interiors. In recent
years, there is growing evidence that their atmospheres might be
enriched with CNO processed material even in the early stages of their
lives, contrary in general to the predictions made by the classical
models of stellar evolution (Schaller et al. \cite{Schaller92}, Chiosi
\cite{Chiosi98}).

Such models predict that only very massive stars, with masses higher
than 60 $M_{\sun}$ on the ZAMS, lose enough mass in the wind to show,
in the main sequence phase, the inner CNO enriched material (Schaller
et al. \cite{Schaller92}, Heger \cite{Heger98}).

For lower mass stars, the exposition of CNO processed material in the
stellar atmosphere only occurs in later phases of the evolution, after
the first dredge-up (Chiosi \cite{Chiosi98}).

However, recent investigations on the effect of stellar rotation on
the structure and evolution of massive stars (Maeder \& Meynet
\cite{Mae&Mey00}, Heger \cite{Heger98}, Heger \& Langer
\cite{HegyLang00}) show that early atmospheric contamination can occur
by mixing with the stellar interior, this contamination being more
important as both the stellar mass and the initial rotational velocity
increase.

The physical mechanism responsible for the mixing processes in massive
stars may be other than rotation, but {\it rotation appears as a
natural driver for this mixing or, at least, as one of the first
mechanisms whose consequences on mixing have to be explored} (Meynet
\& Maeder \cite{Mey&Mae00}), especially considering the fact that
massive stars are in general quite fast rotators.

Rotationally induced mixing has very important consequences on the
evolution of the star, changing the evolutionary tracks towards higher
luminosities, widening the main sequence in the HR diagram and making
it last longer, and also changing the surface composition of the
star. The effects of the CNO cycle show up in the atmosphere, with He
and N enrichments and C and O depletions (Heger \& Langer
\cite{HegyLang00}, Meynet \& Maeder \cite{Mey&Mae00}).  Like Heger \& 
Langer (\cite{HegyLang00}) show in their fig. 9, He enrichments appear
later in the stellar atmosphere, while CNO abundance changes are already
present.

There is plenty of evidence in the literature of this CNO
contamination in early type stars, one of the most important being the
so--called {\it He discrepancy} (Herrero et al.,
\cite{Herrero92}). They find that a considerable fraction of Galactic
OB stars are overabundant in He.

With regard to CNO abundances, a wealth of work also finds the trend in
these elements abundances produced by the CNO cycle. Gies \& Lambert
(\cite{GyLambert92}) find N (and He) overabundances in a sample of B
supergiants in the Galaxy, while Venn (\cite{Venn95}) finds such trend
in the abundances of CNO for A supergiants, but not in the He abundances,
that she assumes to have solar values. Other more qualitative works, 
studying equivalent
widths, also find this trend in the CNO line intensities of B
supergiant stars (Lennon et al., \cite{Lennon93}, McErlean et al.,
\cite{McErlean99}), but not in the He abundance.

Therefore, no clear correlation between the He overabundance and CNO
contamination is found in the literature,  maybe due to the fact that,
as predicted by rotating evolutionary models, changes in the N/C and 
N/O ratios can appear before any He enrichment.

 As can be seen from the
references given, little work has been done on the calculation of CNO
abundances of O stars.  Only Sch\"onberner et al. (\cite{Schon88})
have studied three stars from O8 to O9 and Pauldrach et
al. (\cite{Paul94}) and Taresch et al.  (\cite{Taresch97}) two
individual Galactic early O stars, $\zeta$\,Pup (O4\,If) and
HD\,93129A (O3\,If$^{*}$) respectively.

In the very numerous works on abundances of Galactic B stars (Kilian
et al., \cite{Kil94}, Vrancken et al., \cite{Vran00}, Smartt et al.,
\cite{SmayR97}, Rolleston et al., \cite{Rolles00}, Cunha \& Lambert,
\cite{CunhayL94}, Daflon et al., \cite{Daflon99}, Gummersbach et al.,
\cite{Gumm98}) there is only a marginal presence of late O stars in
the samples studied. Here is where we make our contribution, studying
the CNO abundances of O type stars in order to compare them with the
He abundance.

Of course, abundances derived from spectral analysis are dependent on
the model assumptions.  For example, as demonstrated by Villamariz
(\cite{Villamariz01}) the He abundances obtained for OB stars from
NLTE plane-parallel and hydrostatic model atmospheres are lowered by
$\sim$20 \% (in the He abundance by number relative to the total H+He,
$\epsilon$(He)) due to the inclusion of {\it line--blocking} in the
spectral synthesis. A similar effect is found by consideration of a
microturbulence of 15 km\,s$^{-1}$ in the synthesis of the H and He
lines~of supergiant stars (Villamariz \& Herrero \cite{Villamariz00}),
as found earlier for B supergiants by McErlean et
al. (\cite{McErlean98}) and Smith \& Howarth
(\cite{Smith&Howarth98}). The combined result is that the {\it He
discrepancy} is considerably reduced and it even disappears for some
stars. However, it still remains for an important fraction of Galactic
OB stars.

On the other hand, analyses with the unified code of Santolaya-Rey et
al. (\cite{Santolaya97}), which accounts for the stellar wind and
considers spherical extension of the atmosphere, show no systematic
changes in the He abundance (see Herrero et al.
\cite{Herrero00}). Indeed, for 2 of the 7 stars analysed in this work,
higher He abundances are found (see also Herrero et
al. \cite{Herrero95}).

In this work we determine the CNO abundances of four O9 stars with
known He abundances, as a first step in the investigation of the
possible link between rotation and surface CNO contamination in the
whole range of OB stars. In sect. \ref{sample} we present the objects
selected for this study and the equivalent width measurements. In
sect. \ref{method} we present the technical details of our
methodology, and in sect. \ref{errors} we make a careful study of the
uncertainties in our abundance determinations. Sect. \ref{absolute}
contains the absolute abundances of the three He normal stars that we
analyse, and sect. \ref{relatives} their relative values. Sect. 
\ref{rotator} is devoted to the analysis of the fast rotator 
\object{HD\,191423} and finally the discussion and conclussions of
our work are given in sects. \ref{discuss} and \ref{conclude}
respectively.

\section{The programme stars}\label{sample}
\vspace{0.3cm}

We have studied four O9 stars, two dwarfs and one supergiant with
normal He abundances, and one rapidly rotating giant with an important
He overabundance ($\epsilon$(He)=0.20, Herrero et
al. \cite{Herrero92}).

Our intention is to make a line by line differential analysis within
the four stars, as for example Monteverde et al. (\cite{Ilu00}) and
Smartt et al. (\cite{Smartt96}, \cite{Smartt97}).  The determination
of stellar chemical compositions is known to be dependent on the
details of the methodology used.  That is why it must be done
differentially within stars of similar physical conditions. In this
way, systematic errors in the line abundances, due to the
simplifications adopted in the model atoms for example, can be avoided.

\begin{table}[!h]
\label{stars}
\caption[ ]{The programme stars and their stellar parameters (see text for details). $T_{\rm eff}$ values are given in thousands of Kelvin, 
surface gravities in c.g.s units and the He abundance is the number 
density of He atoms relative to the total number of H+He atoms. Velocities are given in km\,s$^{-1}$. The uncertainties in our parameters are the following: $\Delta T_{\rm eff}$=1\,000 K, $\Delta \log g$=0.10 dex, $\Delta \epsilon$(He)= 25\% and $\Delta \xi$=10 km\,s$^{-1}$} 
\begin{flushleft}
\begin{tabular}{lccccc} 
\hline
~~~~~~Star & v{\thinspace}sin{\thinspace}$i$ & $T_{\rm eff}$ & $\log g$ & $\epsilon$(He) & $\xi$ \\ 
\hline
\object{HD\,214680}, O9 V & 50 & 37.5  & 4.00 & 0.10 & 0 \\ 
\object{HD\,34078}, O9.5 V & 40 & 36.5 & 4.05 & 0.09 & 0 \\ 
\object{HD\,209975}, O9.5 Ib & 100 & 34.0 & 3.20 & 0.08 & 0 \\
                             &     & 35.0 & 3.25 & 0.06 & 20 \\ 
\object{HD\,191423}, O9 III:n & 450 &34.0 & 3.40 & 0.20 & 0 \\ 
                              &     &35.0 & 3.40 & 0.12 & 20 \\ 
\hline
\end{tabular}
\end{flushleft}
\end{table}

\begin{table*}[!t]
\label{eqwis1}

\caption[ ]{Equivalent widths (in m{\AA}) of the C lines identified in
the three stars with normal He abundance (see text for details). Notes
for each line in the third column of each star mean: b: maybe blended,
di: doubtful identification, or clearly blended and not used, e: excluded 
from the final abundance determination, n: narrow, with fwhm considerably 
lower than the rest of the metallic lines.  Equivalent widths quoted as 
-- mean that the line is not present or too weak.}
\begin{center}
\begin{tabular}{lcccccccccc}
\hline
Line &  Multi- & \multicolumn{3}{c}{\object{HD\,214680}} & \multicolumn{3}{c}{\object{HD\,34078}} & \multicolumn{3}{c}{\object{HD\,209975}}  \\
     &  plet &  EW &  $\Delta$EW &Notes &  EW &  $\Delta$EW &Notes &  EW &  $\Delta$EW & Notes \\
\hline
\ion{C}{iii}\,4056 & 24 & 34  &  8 & --  & --  & -- & --  &  -- & --& --  \\
\ion{C}{iii}\,4152 & 21 & 121 & 15 & b,e & 137 & 11 & b,e &  40 & 9 & b,e \\
~~~~+\ion{O}{ii} & \\
\ion{C}{iii}\,4156 & 21 & 64  &  8 & --  & 64  &  6 & --  &  26 & 9 & -- \\
~~~~+\ion{O}{ii} & \\
\ion{C}{iii}\,4162 & 21 & 64  &  7 & --  & 82  &  8 & --  &  34 & 11&--  \\
\ion{C}{iii}\,4186 & 18 & 99  &  7 & --  & 122 &  9 &di &  98 & 12&b,di \\
\ion{C}{iii}\,4325 & 7  & 33 &  6 & --  & 51  &  6 & --  &  11 & 6 & di,e \\
~~~~+\ion{C}{ii}+\ion{O}{ii}+\ion{N}{iii}& \\
\ion{C}{iii}\,4647 & 1  & 268 &  9 & --  & 267 &  7 & --  &  -- & --&--  \\
\ion{C}{iii}\,4665 & 5  & 55  &  9 & --  & 51  &  6 &--   &  35 & 9 & -- \\
~~~~+Si{\sc iii} & \\
\ion{C}{iii}\,4673 & 5  & 41  &  8 & b,e & 40  &  6 & e &  -- & --&-- \\
~~~~+\ion{O}{ii} & \\
\hline
\end{tabular}
\end{center}
\end{table*}

The four stars have been observed by our group during different
observing runs. However, the observing conditions, spectral ranges,
resolutions and S/N values are very similar in all
cases. \object{HD\,209975}, \object{HD\,34078} and \object{HD\,191423}
were observed with the 2.5m Isaac Newton Telescope in 1989 using the
IDS spectrograph with the H2400B grating, resulting in a spectral
resolution of 0.6 \AA~(see Herrero et al. \cite{Herrero92}), whereas
\object{HD\,214680} was observed with the same telescope and
configuration in a later run in 1992 (see Herrero et
al. \cite{Herrero00}). The S/N ratios range from 200 to 300 in the H$_{\rm
\beta}$ region.

The stars were analyzed by Herrero et al. (\cite{Herrero92}) by means
of NLTE plane-parallel and hydrostatic H/He model atmospheres. Here
however we present new parameters with {\it line-blocking} included in
the line formation calculations, obtained in Villamariz
(\cite{Villamariz01}).

Table 1 lists the programme stars together with their
parameters. The parameters of HD\,209975 and HD\,191423 are slightly
changed when the microturbulence found from the CNO analysis is
applied to the H/He analysis, see sections \ref{newgiant} and
\ref{newrotora}.

\subsection{Equivalent width measurements}

We have measured equivalent widths of the CNO lines listed in tables
\ref{eqwis1} and 3 for the three stars with normal He
abundance. The metallic features of the spectrum of the fast rotator
\object{HD\,191423} (see sect. \ref{rotator}) are very weak and
blended because of the high rotational broadening, and for this star
equivalent widths of isolated lines cannot be measured.

When identifying lines in stellar spectra, it is often not easy to
know which blends are contributing to the spectrum at any given
wavelength. To deal with this difficulty, we have considered all possible 
contributions to each wavelength listed in
the VALD database (Kupka et al. \cite{Kupka99}, Ryabchikova et
al. \cite{Ryab99}) and also in the classical book of Striganov \&
Sventitsky \cite{S&S68}, within $\pm$2 \AA~each side of 
the line. In tables \ref{eqwis1} and 3
line identifications (in the first column) are given with all
possible contributions.

The identification of each spectral line is based on its Doppler shift
and on its intensity compared to that of the other members of the
multiplet, i.e., we only assume the identification of a line
when its Doppler shift is consistent with the rest of the identified
lines and its relative intensity compares well to
the remaining lines of the multiplet (see tables \ref{eqwis1} and
3).

The approximate gaussian shape of the line gives an idea of the
possibility of line blending, and we finally decide about the
reliability of each line a posteriori, when all the line abundances
for each ion are calculated (see fig. \ref{abeqwdiag}).

To measure the equivalent widths we use our own software developed in
IDL. A least squares profile fitting procedure was used, with gaussian
profiles fitting the line and polynomials of degree one or two to 
fit the local continuum. 

Errors in the measurements due to the uncertainty in the position of
the local continuum (estimated as $\pm$ 1/S/N) are quoted
together with the equivalent widths. We have found that this is the
most important source of uncertainty in the given values, among those
coming from the use of gaussian fits,
or from the fitting of the continuum.

We have only considered lines of C~{\sc III}, N~{\sc III} and O~{\sc II} for
the abundance
determination, because they are the most common ions of each element
in our O9 spectra.

\begin{table*}[!hb]
\label{eqwis2}
\caption[ ]{Equivalent widths (in m{\AA}) of the N and O lines identified in the three stars with normal He abundance. Notes like in table 2.}
\begin{center}
\begin{tabular}{lcccccccccc}
\hline
Line & Mult. & \multicolumn{3}{c}{\object{HD\,214680}} & \multicolumn{3}{c}{\object{HD\,34078}} & \multicolumn{3}{c}{\object{HD\,209975}}  \\
     &    &  EW &  $\Delta$EW &Notes &  EW &  $\Delta$EW &Notes &  EW &  $\Delta$EW & Notes \\
\hline
\ion{N}{iii}\,4195 & 6  & 23 & 6 & e & --& -- & --  & 25 & 7  & b,e  \\
~~~~+\ion{O}{ii}+\ion{N}{ii} & \\
\ion{N}{iii}\,4215 & 6 	& 16 & 7 & di,e& --& -- & --  & 14  &  8  & e \\
\ion{N}{iii}\,4325 & 10 & 33 & 6 & di,e& 51 & 6 & di,e& 11  &  6  & di,e \\
~~~~+\ion{C}{ii}-{\sc iii}+\ion{O}{ii}& \\
\ion{N}{iii}\,4327 & 10 & 17 & 9 & di,e& 38 & 7 &b,di,e& 15 &  7  & di,e \\
~~~~+\ion{O}{ii}+Si{\sc iv} & \\
\ion{N}{iii}\,4379 & 17 & 105& 9 & --  & 84& 6  & di& 132 &  19 & b,e  \\
~~~~+\ion{C}{iii} & \\ 
\ion{N}{iii}\,4510 & 3 	& 84 & 8 & --  & 62& 6  &--   & 136 &  16 & -- \\
\ion{N}{iii}\,4514 & 3  & 106& 7 & --  &109& 8  & --  & 176 &  16 & b  \\
~~~~+\ion{C}{iii} & \\ 
\ion{N}{iii}\,4523 & 3 	& 37 & 7 & --  & 34& 8  & --  &  59 &  16 & -- \\
\ion{N}{iii}\,4546 & 13 & 17 & 9 & b,e & --& -- & --  &  -- &  -- & -- \\
\ion{N}{iii}\,4634 & 2  & 81 & 11& e & 60& 7  & e & --  &  -- & --  \\

\hline

\ion{O}{ii}\,4072 & 10 & 93 & 9 & --  & 92& 7& --  &  -- &  -- & --  \\
\ion{O}{ii}\,4075 & 10 & 87 & 6 & --  &104& 6& --  & 78  & 12  & di \\
\ion{O}{ii}\,4087 & 48 & 40 & 7 & di& 32& 6& b & --  & --  & b \\
~~~+\ion{N}{ii} & \\   
\ion{O}{ii}\,4132 & 19 & 31 & 9 & b & 57&10& e & 27  &  9  & e  \\
\ion{O}{ii}\,4156 & 19 & 64 & 8 & di,e & 65& 6&di,e & 29  &  9& di,e \\
~~~+\ion{C}{iii} & \\
\ion{O}{ii}\,4189 & 36 & 36 & 7 & --  & 58& 7&di & --  & --  & --  \\
~~~+Fe{\sc iii} & \\
\ion{O}{ii}\,4253 & 101 & 73 & 10& b & 85& 7& --  & --  & --  & --  \\
\ion{O}{ii}\,4282 & 67-54& 31 & 8 &b,di&--&--&b  & 39  & 12  & b  \\
~~~+\ion{C}{iii} & \\
\ion{O}{ii}\,4288 & 54 & 20 & 5 &di & 28& 8&di & --  & --  & --   \\
~~~+\ion{N}{iii} & \\
\ion{O}{ii}\,4294 & 54 	& 35 & 9 & --  & 39& 7& --  &  6  &  5  & di,n\\
\ion{O}{ii}\,4303 & 54 	& 34 & 6 & --  & 61& 7& --  & 12  &  7  & --   \\
\ion{O}{ii}\,4317 & 2   & -- & --  &  -- & 89& 9& --  & --  & --  & --    \\
~~~+\ion{C}{ii} & \\
\ion{O}{ii}\,4319 & 2 	& -- & -- &--  & 74& 7&--   & --  & --  & --   \\
\ion{O}{ii}\,4325 & 2 	& 33 & 6 &di,e &51 & 6&di,e & 11  &  6  & di,e \\
~~~+\ion{C}{ii}-{\sc iii}+\ion{N}{iii} & \\
\ion{O}{ii}\,4349 & 2 	& 54 & 6 & --  &75 & 5& --  & 32  &  6  & --   \\
\ion{O}{ii}\,4366 & 2 	& 65 & 8 & --  &79&  7& --  & 72  & 15  & b  \\
~~~+\ion{C}{iii} & \\
\ion{O}{ii}\,4327 & 41 	& 17 & 9 &di & 38& 7& di& 13  &  7  & di \\
~~~+\ion{N}{iii}+Si{\sc iv} & \\
\ion{O}{ii}\,4369 & 26 & 44 & 9 & b,e & 28& 7& di& --  & --  & -- \\
~~~+\ion{C}{ii} & \\
\ion{O}{ii}\,4414 & 5-60 & 47 & 9 & --  & 67& 7& --  & 25  &  9  & b \\
~~~+\ion{C}{ii} & \\
\ion{O}{ii}\,4596 & 15 	& 27 & 6 & --  & 40& 7& --  & 18  &  7  & n,e \\
\ion{O}{ii}\,4602 & 93 & 34 &  9& b & 32& 7& b &  8  &  6  & n  \\
~~~+\ion{N}{ii} & \\
\ion{O}{ii}\,4610 & 93-92& 69 & 12& b &-- &--&--  & 41  &  14 & b  \\
\ion{O}{ii}\,4638 & 1 & 66 & 8 & --  &90 & 9& --  & 19  &  6  & di \\
~~~+Si{\sc iii} & \\
\ion{O}{ii}\,4661 & 1 & 53 & 8 & --  &63 & 5& --  & 24  &  8  & di \\
~~~+Cu{\sc ii}& \\
\ion{O}{ii}\,4676 & 1 	& 46 & 7 & --  & 62& 6&--   & --  & --  & --   \\
\ion{O}{ii}\,4699 & 40-25 & 40 & 6 & --  & 63& 6& --  & --  & --  & --   \\
~~~+\ion{N}{ii} & \\
\ion{O}{ii}\,4705 & 25 	& 46 & 8 & --  & 51& 5& --  & 16  &  8  & --  \\
\ion{O}{ii}\,4943 & 33 	& 21 & 5 & --  & 40& 5& --  & --  & --  & --  \\
\hline
\end{tabular}
\end{center}
\end{table*}

\clearpage

\section{Method of analysis}\label{method}

For the three He normal stars for which equivalent widths of isolated
lines are measurable, we make use of the classical method of
curve of growth to determine the absolute abundances. For each star
and element we proceed as follows. We construct a grid of model
spectra for abundances 10 times, 5 times and 2 times higher and lower
than solar, and for the solar abundance, each for 6 values of
microturbulence: 0, 5, 10, 15, 20 and 25 km\,s$^{-1}$.

\begin{figure}[!ht]
\resizebox{\hsize}{!}{\includegraphics{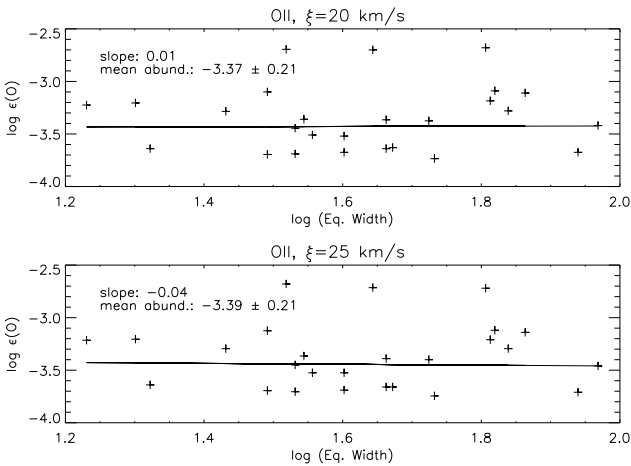}}

\caption[]{Abundance vs. equivalent width for all the \ion{O}{ii}
lines found in the spectrum of \object{HD\,214680}. For a
microturbulence between 20 and 25 km\,s$^{-1}$ the slope reaches 0.
Note how three of the lines (\ion{O}{ii}\,4156, 4325 and
4369) indicate abundances too scattered to be considered.}

\label{abeqwdiag}
\end{figure}

In this way, we construct the curves of growth for each line that
gives the line abundance for each value of microturbulence. With the
abundances of all the lines identified in the stellar spectrum we
obtain the preliminary absolute abundance of the element together with
its particular value of microturbulence (see sect. \ref{micro}). This
value minimizes the dependence of the line abundances with the line
strength in the abundance--equivalent width diagrams. These diagrams
are also a diagnostic tool to check the reliability of the observed
lines for the abundance determination (see fig. \ref{abeqwdiag}).

The model CNO line formation calculations make use of the model
atmosphere and model populations of H and He given by the stellar
parameters $T_{\rm eff}$, $\log g$ and $\epsilon$(He) of each star
(see table \ref{stars}). They are held fixed while the new spectrum is
being calculated, with the codes DETAIL and SURFACE (latest versions,
Butler 1998, private communication).

It is interesting to note that we make a
consistent synthesis for each star and element, for the particular
physical conditions determined by the stellar parameters $T_{\rm
eff}$, $\log g$ and $\epsilon$(He). 
Such a detailed analysis is not common in the literature.
We have studied the methodologies followed in the works with which we will
compare our results: Gies \& Lambert (\cite{GyLambert92}), Cunha \&
Lambert (\cite{CunhayL94}), Sch\"onberner et al. (\cite{Schon88}) and
Daflon et al. (\cite{Daflon99}) and only the latter two make a
consistent synthesis of the metallic spectra.

Gies \& Lambert (\cite{GyLambert92}) and Cunha \& Lambert
(\cite{CunhayL94}) make use of the NLTE metallic abundances of the
Munich group, calculated with the LTE {\it line-blanketed} model
atmospheres of Gold (\cite{Gold84}), while the H/He analysis was made
with the LTE {\it line-blanketed} model atmospheres of Kurucz
(\cite{Kurucz79}), with a more complete {\it blanketing} than that of 
Gold (\cite{Gold84}).

We feel that the additional effort is worthwhile in terms of consistency,
specially for the relative abundances determined, which will hopefully
remain stable whatever physical assumptions are changed in forthcoming
works. Furthermore, Cunha \& Lambert (\cite{CunhayL94}) find that
using the same model atmospheres for the H/He and the
metallic synthesis consistently, apart from leading to different results, 
removes the dependence of the C and N abundances on the stellar temperature.

\subsection{The metallic line opacities}\label{ODFs}

When synthesizing the CNO spectra, we have made extensive use of the line
opacities from the Kurucz opacity
distribution functions (ODF), at solar abundance and 2 kms$^{-1}$
microturbulence (Kurucz \cite{Kurucz96}).

\begin{figure}[!ht]
\resizebox{\hsize}{!}{\includegraphics{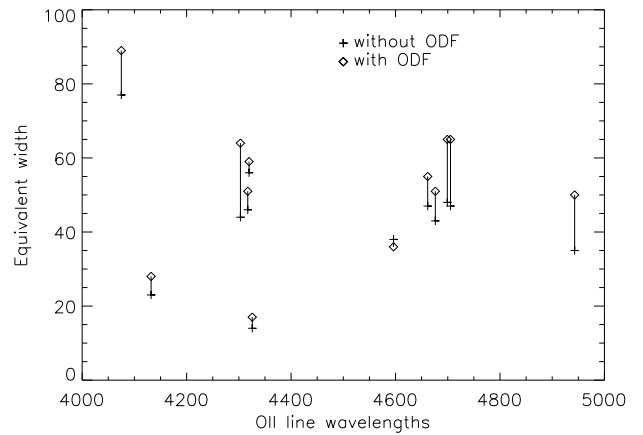}}

\caption[]{Changes in the \ion{O}{ii} model equivalent widths (in
m{\AA}) due to the use of the ODFs of Kurucz (\cite{Kurucz96}) in the
spectral synthesis. Calculations are made at solar abundance and 0
kms$^{-1}$ microturbulence, for the parameters of the star HD\,34078.
Line wavelengths are given in {\AA}.} \label{ODF_O}
\end{figure}

This is to account for the stellar flux blocking due to the metallic
lines, which produces important effects on the synthesized spectra, as
can be seen in fig. \ref{ODF_O} for the particular case of \ion{O}{ii}
in HD\,34078. Therefore, the use of the ODFs will lead to more
appropriate abundances.

For the H/He spectral synthesis we have made use of a different set of
metallic lines, from 228 to 912 \AA~(see Herrero et
al. \cite{Herrero00} for further details) instead of the longer range
(9 -- 100000~\AA)
considered for CNO. However, in the former range we find the strongest
contribution of {\it line-blocking} to the H/He occupation numbers, so
the new opacities considered would yield the same H/He spectra.

\subsection{The model atoms}\label{models}

The model atoms used are basically those described in Becker \& Butler
(\cite{Bec&But88}) for \ion{O}{ii} and in Eber (\cite{Eber87}) for
\ion{C}{iii}. We use a combined \ion{N}{ii}--\ion{N}{iii} model atom
which was constructed using the \ion{N}{ii} model of Becker \& Butler
(\cite{Bec&But89}) and the \ion{N}{iii} model of Butler
(\cite{Butler84}).

\begin{figure}[!ht]
\resizebox{\hsize}{!}{\includegraphics{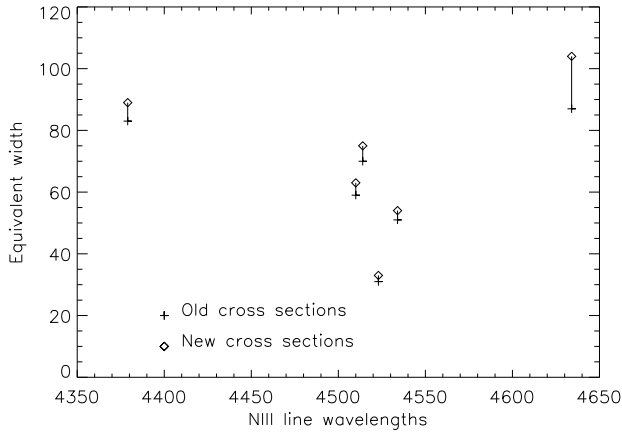}}

\caption[]{Changes in the \ion{N}{iii} model equivalent widths (in
m{\AA}) due to the updated photoionization cross--sections (see text).
Calculations are made at solar abundance and 0 kms$^{-1}$
microturbulence, for the parameters of the star HD\,214680. Line
wavelengths are given in {\AA}.} \label{Ncross}
\end{figure}

We have included a few new transitions in the \ion{O}{ii} and
\ion{C}{iii} model atoms (see table 4) that were observed
in our spectra and are not present in the quoted models. The \ion{N}{ii}
and \ion{N}{iii} photoionization cross sections have been updated
with the Opacity Project atomic data (Cunto et al. \cite{Cunto93}).
The effects of the new cross sections on the synthetic equivalent widths can be seen
in fig. \ref{Ncross}.

\begin{table}[!ht]
\label{newtrans}

\caption[ ]{New transitions included in the atomic models of
\ion{O}{ii} and \ion{C}{iii}. The level energies are taken from the
NIST database (http://physics.nist.gov), in cm$^{-1}$, and the
$\log$\,gf values from VALD. Only the \ion{C}{iii}\,4673 line has been
excluded from the abundance determinations in the whole sample (see text).}

\begin{tabular}{lccc}
\hline
Trans. & E$_l$ & E$_u$ & $\log$\,gf \\
\hline
\ion{C}{iii}\,4056 & 324212.49 & 348859.99 & 0.265 \\
\ion{C}{iii}\,4067 & 322003.68 & 346579.21 &  0.719 \\ 
\ion{C}{iii}\,4068 & 322009.58 & 346579.21 &  0.837 \\
\ion{C}{iii}\,4068 & 322009.58 & 346579.21 & ---0.340 \\
\ion{C}{iii}\,4070 & 322017.97 & 346579.49 &  0.952 \\
\ion{C}{iii}\,4070 & 322017.97 & 346579.21 & ---0.340 \\
\ion{C}{iii}\,4070 & 322017.97 & 346579.21 & ---2.157 \\
\ion{C}{iii}\,4186 & 322702.02 & 346579.31 &  0.918 \\
\ion{C}{iii}\,4325 & 310006.32 & 333118.21 &  0.137 \\ 
\ion{C}{iii}\,4379 & 321411.31 & 344238.68 & ---1.130 \\ 
\ion{C}{iii}\,4379 & 321411.31 & 344236.29 & ---1.255 \\ 
\ion{C}{iii}\,4380 & 321411.31 & 344232.98 & ---2.431 \\ 
\ion{C}{iii}\,4382 & 321426.74 & 344236.29 & ---0.778 \\ 
\ion{C}{iii}\,4383 & 321426.74 & 344232.98 & ---1.255 \\ 
\ion{C}{iii}\,4388 & 321450.05 & 344232.98 & ---0.507 \\ 
\ion{C}{iii}\,4515 & 317794.26 & 339934.72 & ---0.757 \\ 
\ion{C}{iii}\,4515 & 317796.51 & 339934.72 & ---0.280 \\ 
\ion{C}{iii}\,4516 & 317801.30 & 339934.72 & ---0.058 \\ 
\ion{C}{iii}\,4647 & 238213.00 & 259724.30 &  0.007 \\ 
\ion{C}{iii}\,4665 & 308317.29 & 329743.57 &  0.044 \\ 
\ion{C}{iii}\,4673 & 308317.29 & 329706.47 & ---0.434 \\ 
\hline
\ion{O}{ii}\,4253 & 252608.28 & 276109.54 &  0.936 \\ 
\ion{O}{ii}\,4253 & 252608.28 & 276109.46 & ---0.804 \\ 
\ion{O}{ii}\,4253 & 252609.46 & 276109.46 &  0.846 \\ 
\ion{O}{ii}\,4602 & 234402.80 & 256125.78 &  0.510 \\ 
\hline
\end{tabular}
\end{table}

To calculate the Voigt profile of these transitions, we assume that
the collisional broadening is much more important than the radiative
natural broadening and omit the latter. We adopt the formula of Cowley
(\cite{Cow71}) for the collisional parameters.  At
the high temperatures of our models, this is a good assumption because
Doppler broadening dominates the profiles.

During the abundance analysis we have excluded the 
\ion{C}{iii}\,4673 line from the calculation of the C abundance of the three
He normal stars, because it yields considerably higher abundances than
the remainder of the lines. This indicates either that the line is not being well
synthesized, or perhaps that the $\log$\,gf value is underestimated. The
rest of the new transitions give line abundances consistent with the
``old'' ones.

\subsection{Microturbulence}\label{micro}

A wealth of work in the literature (see Vrancken \cite{Vran97a} and
McErlean et al. \cite{McErlean99} for example) show how the value of
microturbulence can depend on which element is considered in a given
atmosphere. The two previous works in particular show how, for B stars, the
value obtained from the \ion{O}{ii} lines is systematically higher
than that obtained from other ions, such as Si{\sc iii}.

We take this into account and make independent
determinations of the abundance together with the microturbulence
value for each element.

In synthesizing the CNO spectra it is necessary to decide how to treat
microturbulence, i. e., whether to include it in both the calculation
of the populations and in the line formation (in both DETAIL and
SURFACE), or only in the line formation (only in SURFACE).

Vrancken  (\cite{Vran97a}) finds that for the high values of
microturbulence of the giants of her sample, the inclusion in both
DETAIL and SURFACE can lead to lower abundances by 0.1-0.2 dex.
As we determine differential abundances, we do not include
microturbulence in DETAIL in the present work.

\begin{table*}[!ht]
\label{errors.C}
\caption[ ]{Line uncertainties in the \ion{C}{iii} lines due to $\Delta T_{\rm eff}$, $\Delta\log g$ and $\Delta\epsilon$(He), in logarithmic scale. ${\Delta\log\epsilon}_{\mathrm i}^{\mathrm {mod}}$ is the total line uncertainty}
\begin{tabular}{cccccccccc}
\hline
Line & \multicolumn{2}{c}{$\Delta T_{\rm eff}$} & \multicolumn{2}{c}{$\Delta\log g$} & \multicolumn{2}{c}{$\Delta\epsilon$(He)} & \multicolumn{2}{c}{${\Delta\log\epsilon}_{\mathrm i}^{\mathrm {mod}}$} & ${\Delta\log\epsilon}_{\mathrm i}^{\mathrm {mod}}$ \\
         & ---1000 K & +1000 K & ---0.10 dex & +0.10 dex & ---0.02 & +0.02 & - & + & \\
\hline
\ion{C}{iii}\,4156 & ---0.07 & 0.09 & 0.16 & ---0.07 & 0.01 &  0.00 & ---0.09& 0.18& $\pm$0.13 \\
\ion{C}{iii}\,4162 & ---0.04 & 0.10 & 0.16 & ---0.06 & 0.01 & ---0.01 & ---0.08& 0.19& $\pm$0.14 \\
\ion{C}{iii}\,4186 & ---0.04 & 0.09 & 0.12 & ---0.05 & 0.00 & ---0.01 & ---0.08& 0.15& $\pm$0.12 \\
\ion{C}{iii}\,4665 & ---0.11 & 0.21 & 0.36 & ---0.11 & 0.02 & ---0.02 & ---0.16& 0.42& $\pm$0.29 \\
\multicolumn{9}{c}{${\Delta\log\epsilon (\mathrm{C})}^{\mathrm {mod}}$} & $\pm$0.19 \\
\hline
\end{tabular}
\end{table*}

\begin{table*}[!ht]
\label{errors.N}
\caption[ ]{Same as table 5 but now for N}
\begin{tabular}{cccccccccc}
\hline
Line & \multicolumn{2}{c}{$\Delta T_{\rm eff}$} & \multicolumn{2}{c}{$\Delta\log g$} & \multicolumn{2}{c}{$\Delta\epsilon$(He)} & \multicolumn{2}{c}{${\Delta\log\epsilon}_{\mathrm i}^{\mathrm {mod}}$} & ${\Delta\log\epsilon}_{\mathrm i}^{\mathrm {mod}}$ \\
         & ---1000 K & +1000 K & ---0.10 dex & +0.10 dex & ---0.02 & +0.02 & - & + & \\
\hline
\ion{N}{iii}\,4510 & 0.18 & ---0.08 & ---0.06 & 0.08 & ---0.00 & 0.02 & ---0.20& 0.10& $\pm$0.15\\
\ion{N}{iii}\,4514 & 0.19 & ---0.10 & ---0.08 & 0.07 & ---0.01 & 0.01 & ---0.20& 0.13& $\pm$0.16\\
\ion{N}{iii}\,4523 & 0.11 & ---0.08 & ---0.05 & 0.04 & ---0.01 & 0.01 & ---0.12& 0.10& $\pm$0.11\\
\multicolumn{9}{c}{${\Delta\log\epsilon (\mathrm{N})}^{\mathrm {mod}}$} & $\pm$0.14 \\
\hline
\end{tabular}
\end{table*}

\section{Abundance uncertainties}\label{errors}

The determination of the abundance uncertainties is as important as
the abundances themselves. 

Our absolute abundances are determined from the mean value of the
individual line abundances (each of them calculated for the mean
microturbulence, see sect. \ref{absolute}) of all the well
behaved lines in the abundance--equivalent width diagrams (see
fig. \ref{abeqwdiag}). In order to consider all the possible sources
of uncertainty in this measurement we proceed following Israelian et
al. (\cite{Isra98}).

The standard deviation of the line abundances, when the number of
lines is large enough and they are weak to strong, will
give us the uncertainty from three different sources: first
from the uncertainty in the position of the continuum, second from the
mean value of microturbulence adopted for the individual abundances,
and third from the limitations of the model atoms, or
in fact of the entire spectral synthesis process.

The other sources of uncertainty come from the uncertainty
in the stellar parameters, $\Delta T_{\rm eff}$, $\Delta\log g$ and
$\Delta\epsilon$(He) ($\pm$1\,000 K, $\pm$0.1 dex and $\pm$25 \%
respectively, see Villamariz \cite{Villamariz01}). 

Villamariz (\cite{Villamariz01}) has shown how H/He model spectra of
lower gravities, those suitable for supergiants, are the most sensitive
to changes in the model calculations. We expect this also to apply to
the metallic spectra, so we estimate the line abundance
uncertainties for our supergiant HD\,209975 and take them as maximum
values of the uncertainty for the whole O9 sample.

The stellar parameters of HD\,209975 are: $T_{\rm eff}$=34\,000~K,
$\log g$=3.20 and $\epsilon$(He)=0.08. For the calculation of the line
uncertainties we proceed as follows. We construct a grid of six model
atmospheres (and H/He populations), putting each parameter in turn to one
of its two extreme values and fixing the other two parameters to the
central values. With each model of the grid we perform all the
abundance calculations as for the central parameters, to obtain new
line abundances (see tables 5 to 7). 

In this way we calculate the line uncertainties due to each
parameter independently, and combine them quadratically to find the final line
uncertainties due to the errors in the stellar parameters
(${\Delta\epsilon}_{\mathrm i}^{\mathrm {mod}}$). In particular, we
have taken the contributions towards higher and lower
abundances separately:

\begin{equation}
{\Delta\epsilon}_{\mathrm i}^{{\mathrm {mod}}\pm}=\sqrt{ {\Delta^2\epsilon}_{\mathrm i}^{T_{\rm eff}\pm} + {\Delta^2\epsilon}_{\mathrm i}^{\log {\mathrm g}\pm} + {\Delta^2\epsilon}_{\mathrm i}^{\epsilon{\rm (He)}\pm} }~,
\end{equation}

and ${\Delta\epsilon}_{\mathrm i}^{\mathrm {mod}}$ is the mean of those two values (see tables 5 to 7).

\begin{table*}[!ht]
\label{errors.O}
\caption[ ]{Same as table 5 but now for O}
\begin{tabular}{cccccccccc}
\hline
Line & \multicolumn{2}{c}{$\Delta T_{\rm eff}$} & \multicolumn{2}{c}{$\Delta\log g$} & \multicolumn{2}{c}{$\Delta\epsilon$(He)} & \multicolumn{2}{c}{${\Delta\log\epsilon}_{\mathrm i}^{\mathrm {mod}}$} & ${\Delta\log\epsilon}_{\mathrm i}^{\mathrm {mod}}$ \\
         & ---1000 K & +1000 K & ---0.10 dex & +0.10 dex & ---0.02 & +0.02 & - & + & \\
\hline
\ion{O}{ii}\,4075 & ---0.13 &  0.21 &  0.24 & ---0.09 &---0.00 & ---0.00 &  ---0.16 & 0.33 & $\pm$0.24\\
\ion{O}{ii}\,4282 & ---0.11 &  0.19 &  0.24 & ---0.10 & 0.01 & ---0.00 &  ---0.15 & 0.30 & $\pm$0.22\\
\ion{O}{ii}\,4294 & ---0.12 &  0.21 &  0.21 & ---0.12 &---0.00 & ---0.00 &  ---0.17 & 0.29 & $\pm$0.23\\
\ion{O}{ii}\,4303 & ---0.09 &  0.21 &  0.24 & ---0.06 & 0.04 & ---0.00 &  ---0.11 & 0.32 & $\pm$0.21\\
\ion{O}{ii}\,4327 & ---0.16 &  0.27 &  0.35 & ---0.13 & 0.03 & ---0.01 &  ---0.21 & 0.45 & $\pm$0.33\\
\ion{O}{ii}\,4349 & ---0.11 &  0.19 &  0.16 & ---0.06 & 0.00 &  0.01 & ---0.13 & 0.24 & $\pm$0.19\\
\ion{O}{ii}\,4366 & ---0.15 &  0.27 &  0.33 & ---0.10 & 0.01 &  0.00 & ---0.18 & 0.42 & $\pm$0.30\\
\ion{O}{ii}\,4414 & ---0.22 &  0.56 &  1.02 & ---0.19 & 0.04 & ---0.01 &  ---0.29 & 1.16 & $\pm$0.72\\
\ion{O}{ii}\,4602 & ---0.10 &  0.16 &  0.20 & ---0.06 &---0.00 & ---0.00 &  ---0.11 & 0.26 & $\pm$0.19\\
\ion{O}{ii}\,4609 & ---0.11 &  0.16 &  0.19 & ---0.09 &---0.00 & ---0.00 &  ---0.15 & 0.25 & $\pm$0.20\\
\ion{O}{ii}\,4638 & ---0.13 &  0.16 &  0.14 & ---0.08 &---0.02 & ---0.00 &  ---0.15 & 0.22 & $\pm$0.18\\
\ion{O}{ii}\,4661 & ---0.13 &  0.19 &  0.16 & ---0.07 &---0.00 & ---0.00 &  ---0.15 & 0.25 & $\pm$0.20\\
\ion{O}{ii}\,4705 & ---0.13 &  0.19 &  0.26 & ---0.10 &---0.00 & ---0.00 &  ---0.16 & 0.32 & $\pm$0.24\\
\multicolumn{9}{c}{${\Delta\log\epsilon (\mathrm{O})}^{\mathrm {mod}}$} & $\pm$0.30 \\
\hline
\end{tabular}
\end{table*}

Again following Israelian et al. (\cite{Isra98}), these individual
uncertainties are used to compute the mean uncertainty for each element
(${\Delta\epsilon (\mathrm{X})}^{\mathrm {mod}}$), which is then
quadratically combined with the standard deviation to give the final
uncertainty in the absolute abundance:

\begin{equation}
\Delta\epsilon(\mathrm{X})=\sqrt{ \sigma^2 + {\Delta^2\epsilon}^{\mathrm {mod}} }~~.
\end{equation}

Remember that while the standard deviation in the line abundances is
individually calculated for each star and element, the uncertainty due
to the model parameters is the same for the whole sample.

\newpage

\section{Absolute abundances of the He normal stars}\label{absolute}

We begin the abundance analysis by determining preliminary absolute
abundances and microturbulence values for each element as explained in
sect. \ref{method}. The line abundance--equivalent width diagrams give
us the set of lines suitable for the abundance determination,
according to the dispersion found in each case.

With the mean value of microturbulence given by the values found for
the three elements, we compute the final line abundances and absolute
abundances, given in table 11. The final absolute abundances
are the mean of the individual values, and its uncertainty is the
combination of the standard deviation and the error coming from the
stellar parameters as explained in sect. \ref{errors}.

Below we give details on the preliminary abundance determination for
each star, and then we give the final absolute abundances for the
three stars in table 11.

\subsection{\object{HD\,214680} (10 Lac), O9 V}

This star belongs to Lac\,OB1, an association at 368 pc from the Sun
according to the Hipparcos data (de Zeeuw et
al. \cite{Zee99}). Kane et al. (\cite{Kane80}) have measured LTE
abundances for two B2 stars of the association, and Gies \& Lambert
(\cite{GyLambert92}) and Sch\"onberner et al. (\cite{Schon88}) for
\object{HD\,214680}.

We only have a few lines for each element in common with Gies \&
Lambert (\cite{GyLambert92}), and the comparison with our equivalent
widths shows that theirs are 5 to 30~\% lower than ours. However,
this difference is what they find when comparing their values to the
literature, and they argue that the reason must be their
underestimation of the far wings of the line when computing the
equivalent width.

Sch\"onberner et al. (\cite{Schon88}) have a total of nine lines of
CNO in common with us, and even though their spectra are photographic
and with a lower S/N, there is good agreement between our respective
values.

\begin{table}[!ht]
\begin{center}
\label{214680.pre}
\caption[ ]{Preliminary absolute abundances for \object{HD\,214680},
O9 V. For each element n is the number of lines used, $\xi$ is the microturbulence and $\log\epsilon(\mathrm X)$ is the logarithmic abundance by number relative to the total H+He. 
The uncertainties given here are $\sigma/\sqrt{\mathrm n}$}
\begin{tabular}{cccccc}
\hline
\multicolumn{2}{c}{C} & \multicolumn{2}{c}{N} & \multicolumn{2}{c}{O}  \\
$\xi$ & n & $\xi$ & n & $\xi$ & n  \\
\multicolumn{2}{c}{$\log\epsilon$(C)} & \multicolumn{2}{c}{$\log\epsilon$(N)} & \multicolumn{2}{c}{$\log\epsilon$(O)} \\
\hline
22.5 & 7 & 10 & 4 & 22.5 & 23 \\
\multicolumn{2}{c}{---3.97$\pm$0.06} & \multicolumn{2}{c}{---4.11$\pm$0.05} & \multicolumn{2}{c}{---3.38$\pm$0.04} \\
\hline
\end{tabular}
\end{center}
\end{table}

The preliminary absolute abundances of CNO for this star are given in
table 8. The microturbulence of 22.5 km\,s$^{-1}$
indicated by the \ion{O}{ii} lines is not a good value. These lines
are too weak to be sensitive to microturbulence (the
maximum equivalent width is only 93 m{\AA}, see table
3), and the scatter in the line abundances is not
minimum for this value of microturbulence, so we do not take it as
being representative.

Note that \ion{O}{ii} in spite of being the most numerous ion in the
spectrum has an uncertainty in the preliminary absolute abundance of
the order of those for C and N, showing how the \ion{O}{ii} line
abundances are  more scattered than for the other two elements.

The high value of microturbulence also found for C is an uncomfortable
result. Kudritzki (\cite{Kud92}) and Lamers \& Achmad
(\cite{LamyAch94}) justify the high values found for supergiants as a
result of the presence of stellar winds, but in the case of this dwarf
star, without evidence in the spectrum of a strong wind, we cannot
attribute this high value of microturbulence to this.

With the microturbulence indicated by N and C we find a mean
microturbulence of 16 $\pm$ 9 km\,s$^{-1}$, with which we compute the
final line and absolute abundances.

\newpage
\subsection{\object{HD\,34078}, O9.5 V}

This star belongs to Ori\,OB1, which is located at about 500 pc from
the Sun (de Zeeuw et al. \cite{Zee99}). Gies \& Bolton
(\cite{GiesyB86}) classify this star as a {\it runaway}, with its 
origin in Ori\,OB1 and an actual peculiar radial
velocity of 50 km\,s$^{-1}$.

Ori\,OB1 is a very well known association, with a wealth of abundance
studies such as those of Kilian (\cite{Kil92}), Cunha \& Lambert
(\cite{CunhayL92}, \cite{CunhayL94}) and Cunha et
al. (\cite{Cunha98}). They all determine abundances of early B dwarf
stars, with which we compare our results.

Gies \& Lambert (\cite{GyLambert92}) have determined abundances for this
star. As for \object{HD\,214680} our equivalent widths are higher than
theirs, due to the reasons stated.

\begin{table}[!ht]
\begin{center}
\label{34078.pre}
\caption[ ]{As in table 8 for \object{HD\,34078}, O9.5 V.}
\begin{tabular}{cccccc}
\hline
\multicolumn{2}{c}{C} & \multicolumn{2}{c}{N} & \multicolumn{2}{c}{O}  \\
$\xi$ & n & $\xi$ & n & $\xi$ & n  \\
\multicolumn{2}{c}{$\log\epsilon$(C)} & \multicolumn{2}{c}{$\log\epsilon$(N)} & \multicolumn{2}{c}{$\log\epsilon$(O)} \\
\hline
12.5 & 6 & 15 & 4 & 17.5 & 23 \\
\multicolumn{2}{c}{---3.82$\pm$0.05} & \multicolumn{2}{c}{---4.18$\pm$0.09} & \multicolumn{2}{c}{---3.37$\pm$0.05} \\
\hline
\end{tabular}
\end{center}
\end{table}

\vspace{-0.25cm}
The preliminary absolute abundances of CNO for this star are given in
table 9. The microturbulence of 17.5 km\,s$^{-1}$
indicated by the \ion{O}{ii} lines is again disregarded for the same
reasons argued for \object{HD\,214680}. And again, the line abundances
for O show more scatter than for C and N.

With the microturbulence indicated by N and C we obtain a mean
microturbulence of 14 $\pm$ 2 km\,s$^{-1}$, with which we compute the
final absolute abundances.

\subsection{\object{HD\,209975}, O9.5 Ib}\label{newgiant}

This star belongs to Cep\,OB2, an association at 615 pc from the Sun
(de Zeeuw et al. \cite{Zee99}). Daflon et al. (\cite{Daflon99}) have
measured abundances for dwarfs from O9 to B2 in this association.

\object{HD\,209975} has a higher rotational velocity than the
two dwarfs studied, and so its lines are more blended and diluted,
leading to higher uncertainties in the equivalent widths and to a
lower number of isolated lines suitable for the abundance analysis
(see tables \ref{eqwis1} and 3).

\begin{table}[!ht]
\begin{center}
\label{209975.pre}
\caption[ ]{As in table 8 for \object{HD\,209975}, O9.5 Ib.}
\begin{tabular}{cccccc}
\hline
\multicolumn{2}{c}{C} & \multicolumn{2}{c}{N} & \multicolumn{2}{c}{O}  \\
$\xi$ & n & $\xi$ & n & $\xi$ & n  \\
\multicolumn{2}{c}{$\log\epsilon$(C)} & \multicolumn{2}{c}{$\log\epsilon$(N)} & \multicolumn{2}{c}{$\log\epsilon$(O)} \\

\hline
12.5 & 4 & 25 & 3 & $>$25 & 13   \\
\multicolumn{2}{c}{---4.03$\pm$0.06} & \multicolumn{2}{c}{---3.97$\pm$0.01} & \multicolumn{2}{c}{$<$---3.63$\pm$0.12} \\
\hline
\end{tabular}
\end{center}
\end{table}

\vspace{-0.25cm}
While the set of lines finally excluded for the abundance
determination in the two dwarfs were coincident for \ion{C}{iii} and
\ion{N}{iii}, and only differ in a single line for \ion{O}{ii},
in the case of the supergiant \object{HD\,209975} we find that the set
of lines that must be excluded according to their dispersion in the
abundance--equivalent width diagrams, are not completely
coincident. To assure consistency in our analysis, we decided to exclude
both sets of lines for every element, so that only those lines with
suitable scatter and also considered in the two dwarfs contribute
to the abundance determination.

This leads to the preliminary abundances in table 10.
The scatter in the \ion{O}{ii} line abundances is
especially severe in this star: compare the
$\sigma/\sqrt{\mathrm n}$ values for the dwarfs in tables
8 and 9 of 0.04 and 0.05, to the 0.12
value for HD\,209975. Again the microturbulence value indicated by O
is not considered to be representative.

With a mean microturbulence of $\xi$=20 $\pm$ 10 km\,s$^{-1}$ we
compute the final absolute abundances for \object{HD\,209975} given in
table 11  ($\log\epsilon_{{\mathrm i}1}$). As seen in Villamariz 
\& Herrero
(\cite{Villamariz00}), the use of a value of $\xi$=15 km \,s$^{-1}$ for
the H/He spectral synthesis of models suitable for supergiants can
lead to changes in the stellar parameters of the order of our standard
error box of 1\,000 K in $T_{\rm eff}$, 0.1 dex in $\log g$ and 25~\% 
in $\epsilon$(He). The changes in
the stellar parameters influence the line abundances
determined as explained in sect~4, so we apply the correspondent corrections to these
abundances due to the value of microturbulence found for this star.

The changes in the parameters are adopted from those found for the O9
supergiant HD\,210809 in Villamariz \& Herrero (\cite{Villamariz00}):
$\Delta T_{\rm eff}$=+1\,000 K, $\Delta \log g$=+0.05 dex and $\Delta
\epsilon$(He)=--- 20 \%. With them we compute the line abundances of
the second column in table 11 for HD\,209975, which are the
final values. As seen in the table these results are
more consistent with the two dwarfs HD\,214680
and HD\,34078.

\begin{table*}[!ht]
\label{abunds}

\caption[ ]{Final line abundances and absolute abundances for the
three He normal stars. Equivalent widths are also given. For
HD\,209975 we give two sets of line abundances:
$\log\epsilon_{{\mathrm i}2}$ are the final values, corrected for the effect
of microturbulence in the stellar parameters (see text).}

\begin{center}
\begin{tabular}{|l c c c c c c c|}
\hline
Line & \multicolumn{2}{ c }{HD\,214680} & \multicolumn{2}{ c }{HD\,34078 } &\multicolumn{3}{ c|}{HD\,209975} \\
     & EW & $\log\epsilon_{\mathrm i}$ & EW & $\log\epsilon_{\mathrm i}$ &  EW & $\log\epsilon_{{\mathrm i}1}$ & $\log\epsilon_{{\mathrm i}2}$  \\
\hline
\ion{C}{iii}\,4056 & 34  & ---4.03 & --  &  --   & -- &  --   &   --  \\
\ion{C}{iii}\,4156 & 64  & ---3.79 & 64  & ---3.78 & 26 & ---4.12 & ---4.07 \\
\ion{C}{iii}\,4162 & 64  & ---3.94 & 82  & ---3.78 & 34 & ---4.14 & ---4.07 \\
\ion{C}{iii}\,4186 & 99  & ---4.11 & 122 & ---3.92 & 98 & ---4.17 & ---4.11 \\
\ion{C}{iii}\,4325 & 33  & ---4.15 & 51  & ---3.95 & -- &  --   &   --  \\
\ion{C}{iii}\,4647 & 268 & ---3.66 & 267 & ---3.63 & -- &  --   &   --  \\
\ion{C}{iii}\,4665 & 55  & ---3.77 & 51  & ---3.86 & 35 & ---3.92 & ---3.77 \\
\hline
$\log\epsilon$(C) & \multicolumn{2}{ c }{---3.89$\pm$0.22} & \multicolumn{2}{ c }{---3.81$\pm$0.16} & \multicolumn{2}{ c}{---4.08$\pm$0.23} & ---3.98 $\pm$0.24 \\
\hline
\ion{N}{iii}\,4379 & 105& ---4.39 & 84& ---4.39 &  -- &  --   &   --  \\
\ion{N}{iii}\,4510 & 84 & ---4.13 & 62& ---4.18 & 136 & ---3.94 & ---4.00 \\
\ion{N}{iii}\,4514 & 106& ---4.11 &109& ---3.91 &  16 & ---3.89 & ---3.98 \\
\ion{N}{iii}\,4523 & 37 & ---4.16 & 34& ---4.05 &  59 & ---3.96 & ---4.04 \\
\hline
$\log\epsilon$(N) & \multicolumn{2}{ c }{---4.19$\pm$0.28} &\multicolumn{2}{ c }{---4.10$\pm$0.28} & \multicolumn{2}{ c}{---3.93$\pm$0.15} & ---4.01$\pm$0.17 \\
\hline
\ion{O}{ii}\,4072 & 93 & ---3.35 & 92 & ---3.56 & -- &  --   &  --   \\
\ion{O}{ii}\,4075 & 87 & ---3.61 &104 & ---3.67 & 78 & ---3.75 & ---3.61 \\
\ion{O}{ii}\,4087 & 40 & ---3.49 & 32 & ---3.76 & -- &  --   &  --   \\
\ion{O}{ii}\,4132 & 31 & ---3.07 & -- &  --   & -- &  --   &  --   \\ 
\ion{O}{ii}\,4189 & 36 & ---3.48 & 58 & ---3.44 & -- &  --   &  --   \\
\ion{O}{ii}\,4253 & 73 & ---3.08 & 85 & ---3.18 & -- &  --   &  --   \\
\ion{O}{ii}\,4282 & 31 & ---3.70 & -- &  --   & 39 & ---3.49 & -3.36 \\
\ion{O}{ii}\,4288 & 20 & ---3.16 & 28 & ---3.15 & -- &  --   &  --   \\
\ion{O}{ii}\,4294 & 35 & ---3.34 & 39 & ---3.42 &  6 & ---4.11 & ---3.99  \\
\ion{O}{ii}\,4303 & 34 & ---3.67 & 61 & ----3.49 & 12 & ---4.12 & ---3.94  \\
\ion{O}{ii}\,4317 & -- &  --   & 89 & ---3.08 & -- &  --   &  --   \\
\ion{O}{ii}\,4319 & -- &  --   & 74 & ---3.29 & -- &  --   &  --   \\
\ion{O}{ii}\,4327 & 17 & ---3.22 & 38 & ---3.03 & 13 & ---3.15 & ---2.97   \\
\ion{O}{ii}\,4349 & 54 & ---3.70 & 75 & ---3.70 & 32 & ---4.15 & ---4.02   \\
\ion{O}{ii}\,4366 & 65 & ---3.14 & 79 & ---3.23 & 72 & ---3.15 & ---2.97     \\
\ion{O}{ii}\,4369 & -- &  --   & 28 & ---3.15 & -- &  --   &  --    \\
\ion{O}{ii}\,4414 & 47 & ---3.60 & 67 & ---3.69 & 25 & ---3.66 & ---3.32\\
\ion{O}{ii}\,4596 & 27 & ---3.25 & 40 & ---3.42 & -- &  --   &   --     \\
\ion{O}{ii}\,4602 & 34 & ---3.42 & 32 & ---3.63 &  8 & ---4.11 & ---3.99    \\
\ion{O}{ii}\,4609 & 69 & ---3.24 & -- &  --   & 41 & ---3.50 & ---3.40     \\
\ion{O}{ii}\,4638 & 66 & ---3.04 & 90 & ---3.04 & 19 & ---3.99 & ---3.90   \\
\ion{O}{ii}\,4661 & 53 & ---3.33 & 63 & ---3.44 & 24 & ---4.01 & ---3.88  \\
\ion{O}{ii}\,4676 & 46 & ---3.33 & 62 & ---3.36 & -- &  --   &   --    \\
\ion{O}{ii}\,4699 & 40 & ---3.65 & 63 & ---3.61 & -- &  --   &   --     \\
\ion{O}{ii}\,4705 & 46 & ---3.60 & 51 & ---3.75 & 16 & ---4.00 & ---3.88    \\
\ion{O}{ii}\,4943 & 21 & ---3.63 & 40 & ---3.56 & -- &  --   &   --     \\
\hline
$\log\epsilon$(O) & \multicolumn{2}{ c }{---3.35$\pm$0.27} & \multicolumn{2}{ c }{---3.36$\pm$0.29} & \multicolumn{2}{ c }{---3.63$\pm$0.52} & ---3.46$\pm$0.48 \\
\hline
\end{tabular}
\end{center}
\end{table*}

\section{Relative abundances of the He normal stars}\label{relatives}

As explained in sect. \ref{sample}, computing differential line
abundances will give us the best estimates of the differences in the
chemical compositions of the stars studied. For stars of similar
parameters, and therefore similar physical conditions, the
systematic errors in the line abundances will be the same for all
stars. The errors will thus be reduced by the use of the differential method.

Among the three He normal stars, the two dwarfs HD\,214680 and
HD\,34078, with projected rotational velocities of 50 and 40
km\,s$^{-1}$ respectively, are assumed to provide the reference values
for the abundances of CNO in unmixed stars. A comparison of their
abundances will give information about the consistency of our method,
which should yield similar values for both, and therefore also about the
scatter that can be expected between stars of similar chemical
compositions. Remember that both stars are in the solar neighbourhood
(de Zeeuw et al. \cite{Zee99}), so they are not expected to have had
different initial chemical compositions.

The supergiant HD\,209975, also from the solar neighbourhood and with
v{\thinspace}sin{\thinspace}$i$= 100 km\,s$^{-1}$, is assumed to provide
the CNO reference abundances for stars of lower gravities. The
differential analysis to the dwarf stars will then tell us whether its
slightly different physical conditions, basically the more extensive
atmosphere with lower surface gravity, introduces any changes in
the reference CNO abundances.

The material of this star should not be mixed. Its He abundance is normal and
its rotational velocity is lower than the 200 km\,s$^{-1}$ considered
as the approximate initial minimum value of the surface equatorial
rotation for efficient mixing (Meynet \& Maeder \cite{Mey&Mae00}). But
we must keep in mind that the 100 km\,s$^{-1}$ is a projected value,
and also that the initial rotational velocities of massive stars
decrease due to the loss of angular momentum in the stellar wind
(Meynet \& Maeder \cite{Mey&Mae00}, Heger \& Langer
\cite{HegyLang00}), so the hypothesis of unmixed material must be
checked.  In adittion, a normal atmospheric He abundance is
compatible with increments in the N/C and N/O ratios in the first stages 
of the main sequence evolution (Heger \& Langer \cite{HegyLang00}).

\begin{table}[!ht]
\begin{center}
\label{rels}
\caption[ ]{Line by line differential abundances of the three He normal stars.}
\begin{tabular}{l c c}
\hline
Line & HD\,34078 & HD\,209975 \\
     &    vs	 &     vs  \\
     & HD\,214680& HD\,214680 \\
\hline
\ion{C}{iii}\,4156  & +0.02 & ---0.28 \\
\ion{C}{iii}\,4162  & +0.16 & ---0.13 \\
\ion{C}{iii}\,4186  & +0.20 & +0.00 \\
\ion{C}{iii}\,4325  & +0.20 & ---   \\
\ion{C}{iii}\,4647  & +0.03 & ---   \\
\ion{C}{iii}\,4665  & ---0.08 & +0.00 \\
\hline
$\Delta \log \epsilon$(C) & +0.10$\pm$0.12 & ---0.09$\pm$0.12 \\

\hline
\ion{N}{iii}\,4379  & +0.00 & ---   \\
\ion{N}{iii}\,4510  & ---0.04 & +0.13 \\
\ion{N}{iii}\,4514  & +0.20 & +0.13 \\
\ion{N}{iii}\,4523  & +0.11 & +0.12 \\
\hline
$\Delta \log \epsilon$(N) & +0.08$\pm$0.12 & +0.13$\pm$0.01 \\

\hline
\ion{O}{ii}\,4072 & ---0.21 & ---   \\
\ion{O}{ii}\,4075 & ---0.05 & ---0.00 \\
\ion{O}{ii}\,4087 & ---0.26 & ---   \\
\ion{O}{ii}\,4189 & +0.04 & ---   \\
\ion{O}{ii}\,4253 & ---0.09 & ---   \\
\ion{O}{ii}\,4282 & ---   & +0.34 \\
\ion{O}{ii}\,4288 & +0.01 & ---   \\
\ion{O}{ii}\,4294 & ---0.07 & ---0.65 \\
\ion{O}{ii}\,4303 & +0.18 & ---0.27 \\
\ion{O}{ii}\,4349 & +0.19 & ---0.32 \\
\ion{O}{ii}\,4366 & +0.00 & +0.17 \\
\ion{O}{ii}\,4327 & ---0.08 & +0.25 \\
\ion{O}{ii}\,4414 & ---0.09 & +0.28 \\
\ion{O}{ii}\,4596 & ---0.17 & ---   \\
\ion{O}{ii}\,4602 & ---0.20 & ---0.57 \\
\ion{O}{ii}\,4609 & ---   & ---0.16 \\
\ion{O}{ii}\,4638 & +0.00 & ---0.86 \\
\ion{O}{ii}\,4661 & ---0.11 & ---0.55 \\
\ion{O}{ii}\,4676 & ---0.03 & ---   \\
\ion{O}{ii}\,4699 & +0.04 & ---   \\
\ion{O}{ii}\,4705 & ---0.15 & ---0.28 \\
\ion{O}{ii}\,4943 & +0.07 & ---   \\
\hline
$\Delta \log \epsilon$(O) & ---0.03$\pm$0.13 & ---0.05$\pm$0.36  \\
\hline
\end{tabular}
\end{center}
\end{table}

In table 12 we give the line by line differential analysis
of the three He normal stars. The final relative abundances are
the simple mean of the individual values, and the standard deviation
its corresponding uncertainty.

This is what we find:

\begin{description}

\item The two dwarf stars have CNO abundances indistinguishable
within the uncertainties, with a scatter in the CN values of 0.1 dex
(with HD\,34078 more abundant than HD\,214680) and the same O
abundance.

\item The supergiant HD\,209975 also has CNO abundances that are
indistinguishable from those of the dwarfs. The scatter in CN is also
0.1 dex and O abundance is again the same. It must be noted however, that
the dispersion of the CN abundances goes in the direction of the CNO
contamination: lower C abundance together with higher N abundance (see
also figs. \ref{NCvrot} and \ref{NOvrot}).

\end{description}

\section{The fast rotator \object{HD\,191423}}\label{rotator}

\begin{figure*}[!ht]
\centering{ 
\includegraphics[width=15cm,angle=90]{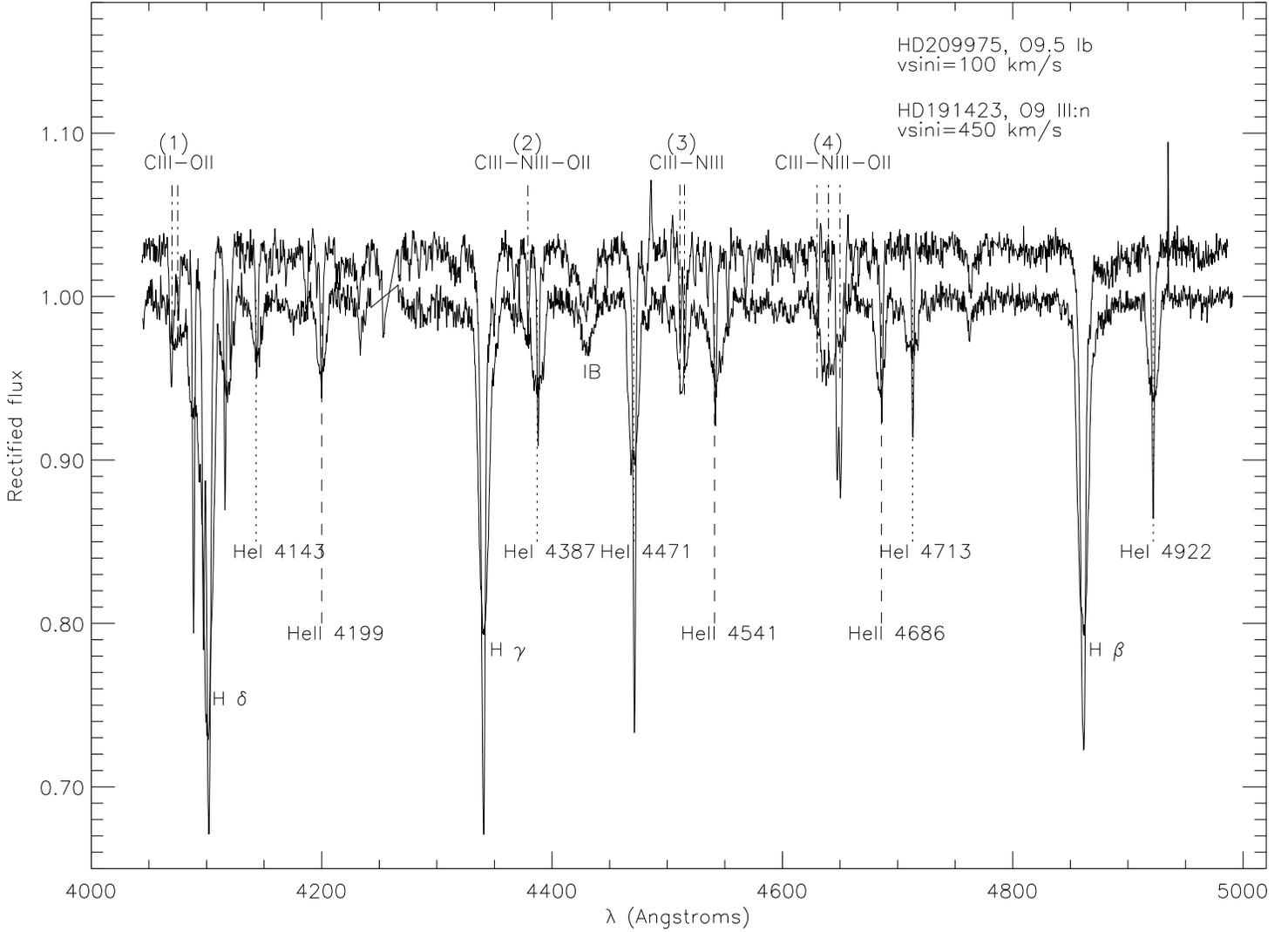}
}
\caption[]{Spectra of HD\,191423 and HD\,209975 (shifted for clarity).
Compare how the lines of the fast rotator are blended while in
HD\,209975 can be still detached.  IB stands for interstellar band.}

\label{191423}
\end{figure*}

This star belongs to the Cyg\,OB8 association, with a distance modulus
of 11.8mag (Humphreys \cite{Hum78}) that situates it at
2.3 kpc from the Sun. \object{HD\,191423} is a very fast rotator,
v$\sin i$=450 km\,s$^{-1}$, with the weak metallic
lines completely blended, so the method used for the other three stars
cannot be used in this case (see fig. \ref{191423}). We perform a
spectral synthesis and as we analyse the H/He spectra, fitting our
model lines to the observed ones to find the stellar parameters
$T_{\rm eff}$, $\log g$ and $\epsilon$(He), we find the best fits
to the observed blends that give us the stellar CNO
abundances. This methodology is widely used in stellar abundance
determinations when the lines available are not isolated (see for
example Pauldrach et al., \cite{Paul94} and Taresch et
al., \cite{Taresch97} for early O stars, and Garc\'\i a-L\'opez et
al., \cite{Ramon95} and Israelian et al., \cite{Isra98} for late type
stars).

We have identified all the metallic spectral features present in the
spectrum of the star. Those that correspond to lines used for the three
He normal stars are now selected for the abundance analysis. See 
fig. \ref{191423} for the four metallic spectral features available:

\begin{description}

\item{ (1): \ion{C}{iii}--\ion{O}{ii}\,4067--82 }
\item{ (2): \ion{C}{iii}--\ion{N}{iii}--\ion{O}{ii}\,4379 }
\item{ (3): \ion{C}{iii}--\ion{N}{iii}\,4505--21 }
\item{ (4): \ion{C}{iii}--\ion{N}{iii}--\ion{O}{ii}\,4620--60 }
\end{description}

We have made a careful search of all the possible contributions to
these blends, as explained in sect. \ref{method} when identifying lines
in the stellar spectra. We find that the dominant contributions to
each blend are those quoted, considering the lines that we have found
in the other three stars in the same ranges, and also the synthetic
spectra (see sect. \ref{CNOrot}). We cannot make a
line by line differential abundance analysis as for the other three
stars, but for the fast rotator we make use of the lines that were
suitable for the abundance determinations in the reference stars.

Blend (4) is in emission in Of stars together with the
\ion{He}{ii}\,4686 line (Walborn \cite{Wal71}). In the case of stars with
weaker winds, like HD\,191423, it can show some filling by the wind
emission even while being in absorption, so we do not expect to obtain good
fits for this blend.

In order to synthesize the complete CNO+H/He spectrum we proceed as
follows. First we calculate individually the populations of each
element, making use of DETAIL and the model atoms quoted in
sect. \ref{models}. Then we use a combined CNO+H/He model atom for
the line formation calculations with SURFACE to reproduce the
blends observed.

 The last step in this process, before the comparison with the stellar
spectrum, is the convolution of the synthetic profiles with a standard 
rotational broadening function (and of course with the instrumental profile,
negligible in this case). This procedure is not very suitable for objects with
such high rotational velocities, where surface deformation and gravity 
darkening become of considerable importance. Further work must be done to 
improve our treatment of rotation.

\subsection{Microturbulence and new stellar parameters}\label{newrotora}

As the entire CNO+H/He spectrum is calculated simultaneously, we have
to make use of one particular value of microturbulence for all the
elements. We consider the mean value found for the supergiant
HD\,209975, $\xi$= 20 km\,s$^{-1}$, because of its similarity with our
fast rotator.

As seen for HD\,209975, this value of the microturbulence for a giant star
like HD\,191423 can lead to changes in the stellar parameters, which
have been determined for $\xi$= 0 km\,s$^{-1}$. Therefore we proceed
to determine new stellar parameters $T_{\rm eff}$, $\log g$ and
$\epsilon$(He) at $\xi$= 20 km\,s$^{-1}$.

\begin{figure}[!ht]
\resizebox{\hsize}{!}{\includegraphics{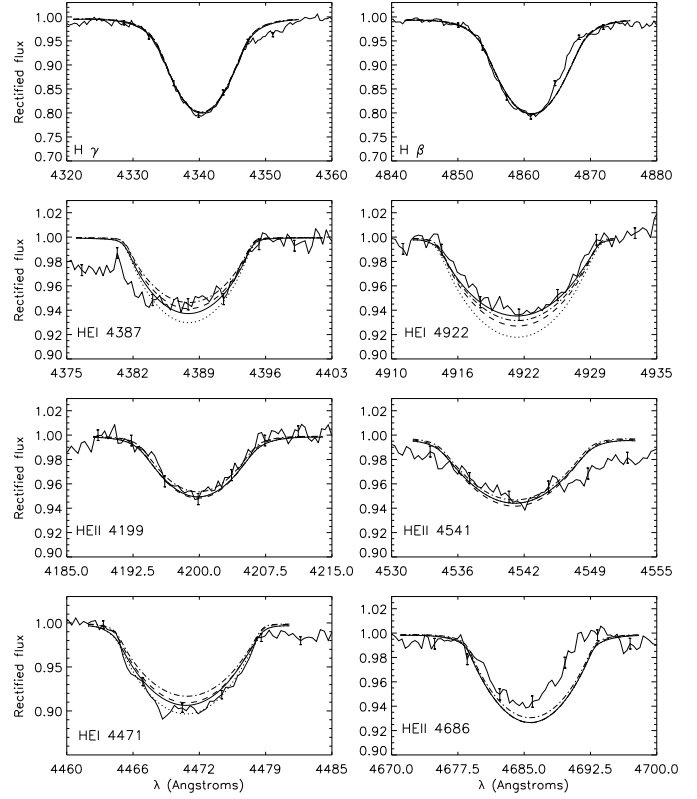}}

\caption[]{Fitting the H/He spectrum of HD\,191423 to determine
$T_{\rm eff}$, $\log g$ and $\epsilon$(He) at $\xi$= 20 km\,s$^{-1}$.
The initial parameters are $T_{\rm eff}$=34\,000 K, $\log g$=3.40,
$\epsilon$(He)=0.20, $\xi$= 0 km\,s$^{-1}$ (solid line), then we plot
the same model spectrum at $\xi$= 20 km\,s$^{-1}$ (dotted), and $T_{\rm
eff}$=35\,000 K, $\log g$=3.40, $\epsilon$(He)=0.15, $\xi$= 20
km\,s$^{-1}$ (dashed). The final parameters are: $T_{\rm eff}$=35\,000~K, 
$\log g$=3.40, $\epsilon$(He)=0.12 and $\xi$= 20 km\,s$^{-1}$
(dash-dotted). The error bars trace the uncertainty in the rectified
flux ($f_\mathrm{rect}$) due to the S/N of the spectrum ($\pm~1/S/N
\cdot 1/f_\mathrm{rect}$)}.

\label{191423pars}
\end{figure}

We start with a model spectrum at $\xi$=20 km\,s$^{-1}$ and the
initial parameters of the star: $T_{\rm eff}$=34\,000 K, $\log g$=3.40
and $\epsilon$(He)=0.20 (see fig. \ref{191423pars}, dotted line) and
we compute new model spectra in order to find the best fit for the
observed H/He lines (for details on the lines selected for the H/He
analysis, see Herrero et al., \cite{Herrero92} and Villamariz,
\cite{Villamariz01}).

\begin{figure}[!ht]
\resizebox{\hsize}{!}{\includegraphics{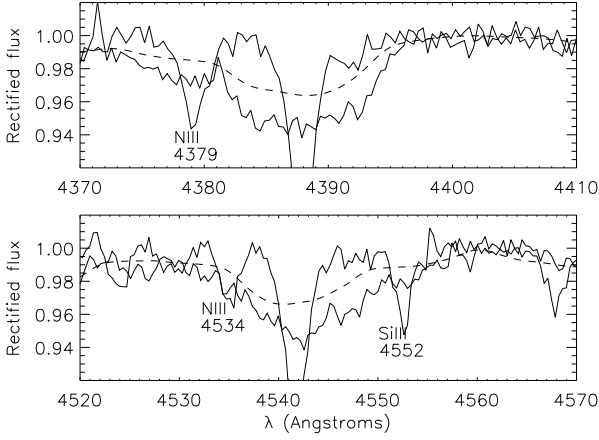}}
\caption[]{Line blending in \ion{He}{i}\,4387 and \ion{He}{ii}\,4541. The spectra of HD\,209975 (solid line, sharp lines), HD\,209975 degraded to 450 km\,s$^{-1}$ (dotted line) and HD\,191423  (solid line, wide features) are shown.}
\label{Heblends}
\end{figure}

In fig. \ref{191423pars} we see the final stage of the fitting,
yielding the new parameters: $T_{\rm eff}$=35\,000 K, $\log g$=3.40,
$\epsilon$(He)=0.12 and $\xi$= 20 km\,s$^{-1}$ (dash-dotted). It is
interesting to note in the figure how the lines \ion{He}{i}\,4387 and
\ion{He}{ii}\,4541 cannot be well reproduced with the H/He model
atom, since they are blended with metallic lines (see
fig. \ref{Heblends}). In the following spectral synthesis we see
how the fits improve with the use of the combined CNO+H/He model atom
(see figures \ref{191423fit0} and \ref{191423fit1}). The blend of
\ion{He}{ii}\,4541 with \ion{Si}{iii}\,4542 is not reproduced by
this combined model atom.

With the new parameters \ion{He}{i}\,4471 shows the {\it dilution
effect} present in the spectra of late O and early B giants and
supergiants (see Voels et al. \cite{Voels89}, McErlean et
al. \cite{McErlean98}, Herrero et al. \cite{Herrero92}), and
\ion{He}{ii}\,4686 shows some filling with wind emission.

On the basis of these new parameters we compute all the CNO+H/He synthetic
spectra.

\subsection{CNO abundances in \object{HD\,191423}, O9 III:n}\label{CNOrot}

To start the abundance analysis of the fast rotator HD\,191423 we
begin with the abundances found for the reference supergiant
HD\,209975:

\begin{description}
\item $\log\epsilon(\mathrm {C})= -3.98 \pm 0.24 \sim \log{\frac{{\epsilon(\mathrm {C})}_{\odot}}{5}}= -4.19$
\item $\log\epsilon(\mathrm {N})= -4.01 \pm 0.17 \sim \log{\epsilon(\mathrm {N})_{\odot}}= -4.07$
\item $\log\epsilon(\mathrm {O})= -3.46 \pm 0.48 \sim \log{\frac{{\epsilon(\mathrm {O})}_{\odot}}{5}}= -3.87$\footnote{In fact this value is closer to $\log{ \frac{{\epsilon(\mathrm {O})}_{\odot}}{2}}= -3.47$, but a transcription error made us start with the quoted value. Nevertheless our results are not dependent on this. Solar abundances are taken from Grevesse et al. (\cite{Grev96})}
\end{description}

\begin{figure}[!ht]
\resizebox{\hsize}{!}{\includegraphics{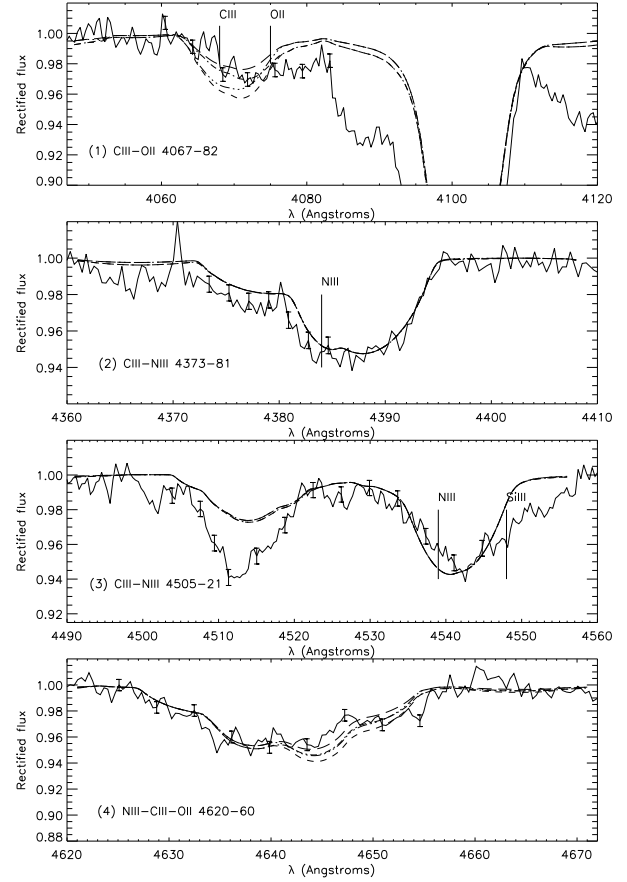}}
\caption[]{Spectral synthesis for HD\,191423, for the model abundances M0 (short dashed), M1 (dash--dotted), M2 (dash--three-dotted), and M3 (long dashed), see text. M1 gives the best fit to blend (1), with $\epsilon$(C)=$\frac{{\epsilon(\mathrm{C})}_{\odot}}{10}$, $\epsilon$(O)=$\frac{{\epsilon(\mathrm{O})}_{\odot}}{5}$ , $\epsilon$(He)=0.12  and $\xi$=20 km\,s$^{-1}$}
\label{191423fit0}
\end{figure}

With the CNO abundances of our standard grid of 10, 5 and 2 times
higher and lower than solar that are closest to the abundances of
HD\,209975 we compute the first model spectrum (model M0, see
fig. \ref{191423fit0}, short--dashed line).

The blend labeled as (1) is the only one that gives us information
about the abundances of O and C. To find a good fit, starting from M0,
we must lower those abundances, which we do in models M1
to M3:

\begin{description}

\item {M0: $\epsilon(\mathrm{C})=\frac{{\epsilon(\mathrm{C})}_{\odot}}{5}$, $\epsilon(\mathrm{N})={\epsilon(\mathrm{N})}_{\odot}$, $\epsilon(\mathrm{O})=\frac{{\epsilon(\mathrm{O})}_{\odot}}{5}$}

\item {M1: $\epsilon(\mathrm{C})=\frac{{\epsilon(\mathrm{C})}_{\odot}}{10}$, $\epsilon(\mathrm{N})={\epsilon(\mathrm{N})}_{\odot}$, $\epsilon(\mathrm{O})=\frac{{\epsilon(\mathrm{O})}_{\odot}}{5}$}

\item {M2: $\epsilon(\mathrm{C})=\frac{{\epsilon(\mathrm{C})}_{\odot}}{5}$, $\epsilon(\mathrm{N})={\epsilon(\mathrm{N})}_{\odot}$, $\epsilon(\mathrm{O})=\frac{{\epsilon(\mathrm{O})}_{\odot}}{10}$}

\item {M3: $\epsilon(\mathrm{C})=\frac{{\epsilon(\mathrm{C})}_{\odot}}{10}$, $\epsilon(\mathrm{N})={\epsilon(\mathrm{N})}_{\odot}$, $\epsilon(\mathrm{O})=\frac{{\epsilon(\mathrm{O})}_{\odot}}{10}$}
\end{description}

Blends (2) and (3) have \ion{O}{ii} components, but as can be seen
in fig. \ref{191423fit0}, they are dominated by the \ion{N}{iii}
component, the only line that remains at a fixed abundance in models M0
to M3.

Looking at the fits in fig. \ref{191423fit0} we obtain the C and O
abundances of HD\,191423:
$\epsilon(\mathrm{C})=\frac{{\epsilon(\mathrm{C})}_{\odot}}{10}$,
$\epsilon(\mathrm{O})=\frac{{\epsilon(\mathrm{O})}_{\odot}}{5}$ (model
M1, dash--dotted), which are held fixed while we search for the N
abundance, with models M4 and M5:

\begin{description}

\item {M4: $\epsilon(\mathrm{C})=\frac{{\epsilon(\mathrm{C})}_{\odot}}{10}$, $\epsilon(\mathrm{N})=2.5\cdot{\epsilon(\mathrm{N})}_{\odot}$, $\epsilon(\mathrm{O})=\frac{{\epsilon(\mathrm{O})}_{\odot}}{5}$}

\item {M5: $\epsilon(\mathrm{C})=\frac{{\epsilon(\mathrm{C})}_{\odot}}{10}$, $\epsilon(\mathrm{N})=5\cdot{\epsilon(\mathrm{N})}_{\odot}$, $\epsilon(\mathrm{O})=\frac{{\epsilon(\mathrm{O})}_{\odot}}{5}$}

\end{description}

\begin{figure}[!ht]
\resizebox{\hsize}{!}{\includegraphics{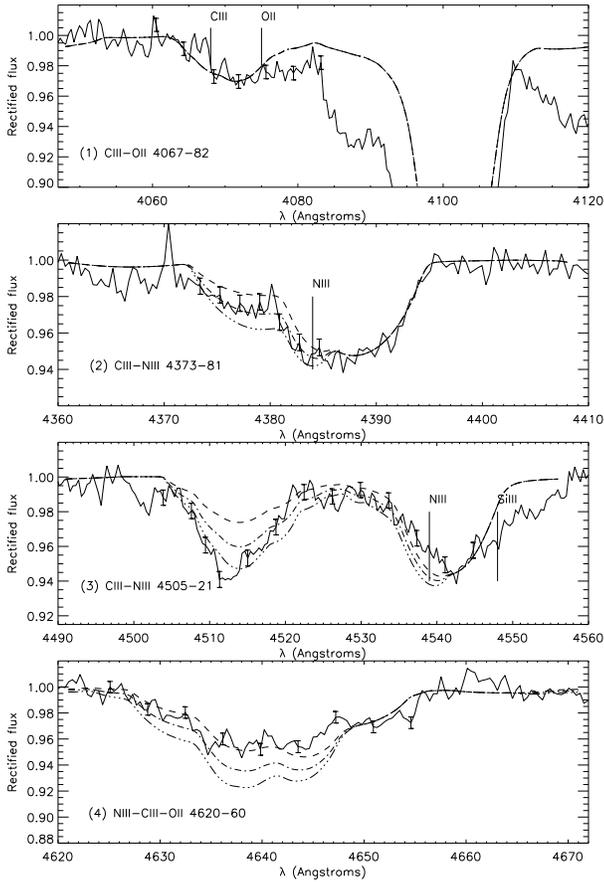}}
\caption[]{Spectral synthesis for HD\,191423, for the model abundances M1 (dashed), M4 (dash--dotted) and M5 (dash--three-dotted) which only differ in the N abundance. See how blend (2) and (3) indicate different N abundances (see text)}
\label{191423fit1}
\end{figure}

In fig. \ref{191423fit1} we see that it is not possible to find a good
fit for both blends (2) and (3) with a single value of the N
abundance. While blend (2) and the \ion{N}{iii} line embedded in
\ion{He}{i}\,4387 are well reproduced for
$\epsilon(\mathrm{N})=2.5\cdot{\epsilon(\mathrm{N})}_{\odot}$, blend
(3) is better reproduced by a higher abundance,
$\epsilon(\mathrm{N})=5\cdot{\epsilon(\mathrm{N})}_{\odot}$. Furthermore,
the \ion{N}{iii} line embedded in \ion{He}{ii}\,4541 also gives
contradictory information, because it is too strong with the three N
abundances considered. We then find that the N abundance of this star
may be between 2.5 and 5 ${\epsilon(\mathrm{N})}_{\odot}$.

Blend (4) shows evidence of wind emission, as with \ion{He}{ii}\,4686.

Adopting the abundance uncertainties of HD\,209975 we give the
absolute CNO abundances of the fast rotator HD\,191423:

\begin{description}
\item $\log \epsilon$(C)=---4.48$\pm$0.24
\item $\log \epsilon$(N) $\in$[---3.70,---3.40]$\pm$0.17
\item $\log \epsilon$(O)=---3.82$\pm$0.48
\end{description}

From now on we consider the most conservative value for the N
abundance, the lower value, to be the N abundance of HD\,191423.


\section{Discussion}\label{discuss}

\begin{figure}[!ht]
\resizebox{\hsize}{!}{\includegraphics{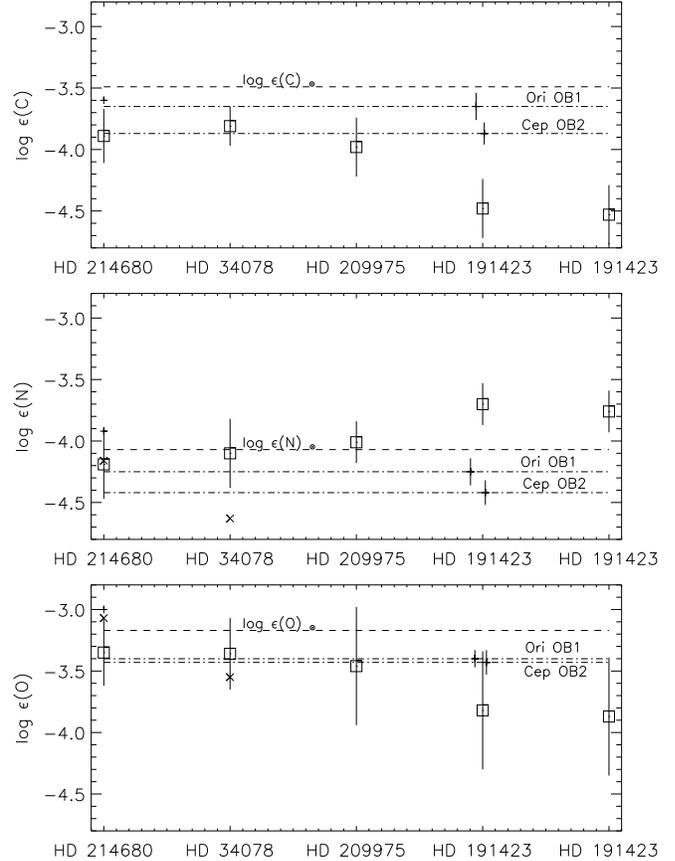}} \caption[]{CNO
abundances of our sample stars compared to the values of Gies \&
Lambert (\cite{GyLambert92}, crosses) for HD\,214680 and HD\,34078, and
of Sch\"onberner et al. (\cite{Schon88}, plus signs) for
HD\,214680. Also the mean abundances for early B dwarfs in Ori\,OB1
(Cunha \& Lambert \cite{CunhayL94}) and in Cep\,OB2 (Daflon et
al. \cite{Daflon99}) are plotted (dash--dotted lines with their error
bars). Solar abundances are also shown for reference. For HD\,191423
we give first the absolute abundances and then the ones corrected for
Galactic abundance gradients (see text).}  \label{CNOall} \end{figure}

In fig. \ref{CNOall} and in table 13 we compare our results
with those of Gies \& Lambert (\cite{GyLambert92}) and Sch\"onberner
et al. (\cite{Schon88}) for HD\,214680, with Gies \& Lambert
(\cite{GyLambert92}) for HD\,34078, and with Cunha \& Lambert
(\cite{CunhayL94}) and Daflon et al. (\cite{Daflon99}) for the
associations Ori\,OB1 and Cep\,OB2 respectively. In Cunha \& Lambert
(\cite{CunhayL92}, \cite{CunhayL94}) it is shown how the two youngest
subgroups of Ori\,OB1 (Ic and Id) are enriched in O and Si with
respect to the other two older subgroups (Ia and Ib). We compare
our results with the mean abundances of the older, non-enriched
stars. Also the solar abundances of Grevesse et al. (\cite{Grev96})
are quoted.

\begin{table*}[!ht]
\label{CNO.abs} \caption[ ]{Absolute CNO abundances of our four O9
stars compared to the values in the literature of Gies \& Lambert
(\cite{GyLambert92}), Sch\"onberner et al. (\cite{Schon88}), Cunha \&
Lambert (\cite{CunhayL94}) and Daflon et al. (\cite{Daflon99}). For 
HD\,191423 we give the abundances corrected for the Galactic gradients
of Rolleston et al. (\cite{Rolles00}) and Gummersbach et al. 
(\cite{Gumm98}), see text.}

\begin{center}
\begin{tabular}{ l c c c c}
\hline
           & $\log\epsilon$(C) & $\log\epsilon$(N) & $\log\epsilon$(O) & Ref.\\
\hline
HD\,214680 & --3.89$\pm$0.22 & --4.19$\pm$0.28 & --3.35$\pm$0.27 & This work\\
O9 V       &                 & --4.16$\pm$0.31 & --3.07$\pm$0.29 & G\&L \\
	   & --3.60$\pm$0.20 & --3.92$\pm$0.20 & --3.00$\pm$0.20 & Sch\"on.\\
\hline
\noalign{\smallskip}
HD\,34078  & --3.81$\pm$0.16 & --4.10$\pm$0.28 & --3.36$\pm$0.29 & This work\\
O9.5 V     &                 & --4.63$\pm$0.09 & --3.55$\pm$0.25 & G\&L\\
\noalign{\smallskip}
Early B V stars & --3.65$\pm$0.11 & --4.25$\pm$0.11 & --3.40$\pm$0.07 & C\&L \\
in Ori\,OB1 &		     & 		       &		 &      \\  
\hline
\noalign{\smallskip}
HD\,209975 & --3.98$\pm$0.24 & --4.01$\pm$0.17 & --3.46$\pm$0.48 & This work\\
O9.5 Ib    &		     & 		       &		 &      \\ 
\noalign{\smallskip}
Early B V stars & --3.87$\pm$0.09 & --4.42$\pm$0.10 & --3.43$\pm$0.10 & Daflon \\
in Cep\,OB2 &		     & 		       &		 &      \\
\hline
\noalign{\smallskip}
HD\,191423 & --4.48$\pm$0.24 & --3.70$\pm$0.17 & --3.82$\pm$0.48 & This work\\
O9 III:n   & --4.50$\pm$0.24 & --3.76$\pm$0.17 & --3.87$\pm$0.48 & Gumm. \\
           & --4.53$\pm$0.24 & --3.76$\pm$0.17 & --3.87$\pm$0.48 & Roll. \\
\hline
\noalign{\smallskip}
Solar abundances & --3.49    &     --4.07      &     --3.17     &Grevesse\\
\hline
\end{tabular}
\end{center}
\end{table*}

Due to the strong dependence of the abundance determinations on the
particular methodologies followed (see our
discussion on this in sects. \ref{method} and \ref{relatives}), the
direct comparison of absolute abundances obtained in different works
must be done with care, considering a possible ``methodological
bias''.

For the two dwarfs we see that our results are more consistent with
each other than
those of Gies \& Lambert (\cite{GyLambert92}), which show considerable
differences in the N and O abundances of these two similar stars (see
fig. \ref{CNOall}, crosses). They do not determine the C
abundances. Sch\"onberner et al. (\cite{Schon88}) find CNO abundances
of HD\,214680 to be higher than ours by 0.3 dex (see fig. \ref{CNOall}, plus
signs), an effect that can be explained by the ODFs used in our work,
which as seen in sect. \ref{ODFs}, lead to lower abundances.

The comparison with Cunha \& Lambert (\cite{CunhayL94}) and Daflon et
al. (\cite{Daflon99}) shows good agreement considering the
uncertainties, and it is remarkable that in both their works and
our dwarf stars, the C and N abundances are more dispersed than the O
abundance, which is practically the same in all cases.

The CNO abundances that we obtain for our two reference dwarf stars
are lower than the solar values by 0.1--0.4 dex, also in good
agreement with the abundances of other unmixed early B dwarfs in the
solar neighbourhood (Kilian \cite{Kil92}, \cite{Kil94}, Vrancken
\cite{Vran97a}). The results therefore support the hypothesis that the Sun is
enriched in these elements relative to its neighbour early type stars.

The CNO abundances of our supergiant HD\,209975 are different from
those of the two reference dwarfs by $\sim$0.1 dex in the direction of
CNO contamination: lower C and O and higher N abundances. However,
these small differences are within our uncertainties (also in the
relative values, see tab. \ref{rels}) , so we believe that the 
differences can be due to the
different physical conditions of the star relative to the dwarfs, as
we indicated in sect. \ref{relatives}, and not to real different
chemical compositions (remember that it also formed in the solar
neighbourhood).

This star gives the reference CNO abundances of unmixed
O9 giants and supergiants, but we will have to study more low gravity
stars in order to establish these reference abundances
reliably.

The CNO absolute abundances of our fast rotator HD\,191423 could be
showing the effect of possible internal mixing together with the
different original composition of the cloud from which it was
formed. It belongs to the Cyg\,OB8 association, that according to
Melnick \& Efremov (\cite{Mel&Efre95}) forms a group with the
associations Cyg\,OB1 and Cyg\,OB9. Adopting for our star the
galactocentric distance of Cyg\,OB1 (R$_{\mathrm GC}$=7.8 kpc,
Massey et al. \cite{Mass95}), we see that it is closer to
the galactic center than the Sun, at 8.5 kpc (Gummersbach et
al. \cite{Gumm98}). 

Considering the Galactic abundance gradients found
by several authors such as Smartt \& Rolleston (\cite{SmayR97}),
Gummersbach et al. (\cite{Gumm98}) and Rolleston et al. 
(\cite{Rolles00}),
we must correct for the different initial compositions before
comparing its abundances with those of our reference stars.

We do so with the most recent gradients of Rolleston et al. 
(\cite{Rolles00}):
\begin{description}
\item $\Delta$C= ---0.07 dex/kpc $\times$ (7.8--8.5)kpc = + 0.05 dex
\item $\Delta$N= ---0.09 dex/kpc $\times$ 0.7 kpc = + 0.06 dex
\item $\Delta$O= ---0.067 dex/kpc $\times$ 0.7 kpc = + 0.05 dex
\end{description}      

\noindent that yield the second set of abundances represented in
fig. \ref{CNOall}. We see that the corrections are small and tend
to reduce the N overabundance and to increase the C and O 
underabundances. Anyway the trend obtained in the CNO abundances 
of this fast rotator is in the direction of CNO contamination.

Massive stellar evolution with rotation predicts that the mixing
is more efficient for higher mass stars and for higher initial 
rotational velocities, with the surface CNO contamination growing
as the star ages. Another prediction is that the
rotation of the star decreases with time, due to the loss of angular
momentum in the stellar wind (Meynet \& Maeder \cite{Mey&Mae00}, Heger 
\& Langer \cite{HegyLang00}).

The actual spectroscopic masses of our sample stars are the
following:
\begin{description}
\item HD\,214680, 18 $M_{\sun}$
\item HD\,34078, 21 $M_{\sun}$
\item HD\,209975, 20 $M_{\sun}$
\item HD\,191423, 29 $M_{\sun}$
\end{description}

\begin{figure}[!ht]
\resizebox{\hsize}{!}{\includegraphics{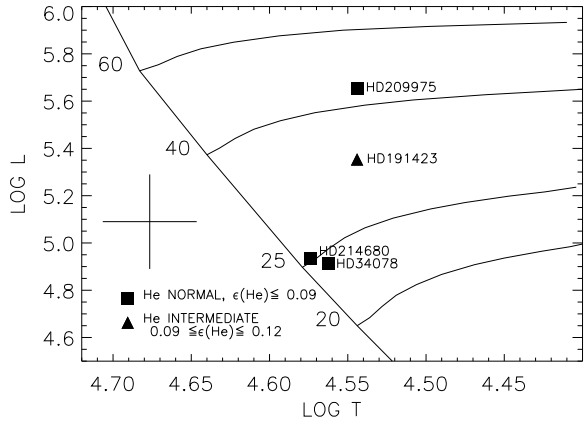}}
\caption[]{HR diagram for the O9 stars studied. Classical main sequence
evolutionary tracks of Schaller et al. (\cite{Schaller92}) are shown 
for the quoted initial masses.}
\label{hrd}
\end{figure}

\noindent and in fig. \ref{hrd} we can see in the HR diagram that our 
two dwarf stars are certainly young, while the supergiant HD\,209975 
and the fast rotator are more evolved.

Therefore, the surface CNO contamination found in HD\,191423 is in
agreement with the evolutionary predictions: the star started its 
main sequence lifetime with a very high initial rotational
velocity and it is now showing the corresponding surface CNO
contamination.

\begin{figure}[!ht]
\resizebox{\hsize}{!}{\includegraphics{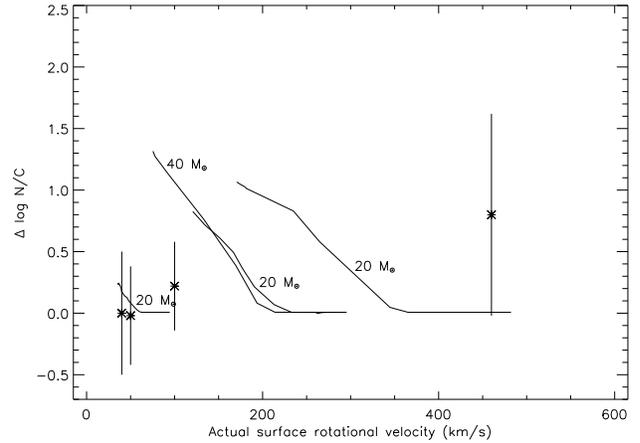}}
\caption[]{Evolution of the N/C ratio during the main sequence as 
a function of the equatorial surface rotational velocity, for models
of the quoted initial masses. The observed data are for the projected 
rotational velocities, and therefore they could all be  right shifted 
to the non-projected values. Uncertainties in v{\thinspace}sin{\thinspace}$i$ range from 10 to 20 km\,s$^{-1}$}
\label{NCvrot}
\end{figure}

\begin{figure}[!ht]
\resizebox{\hsize}{!}{\includegraphics{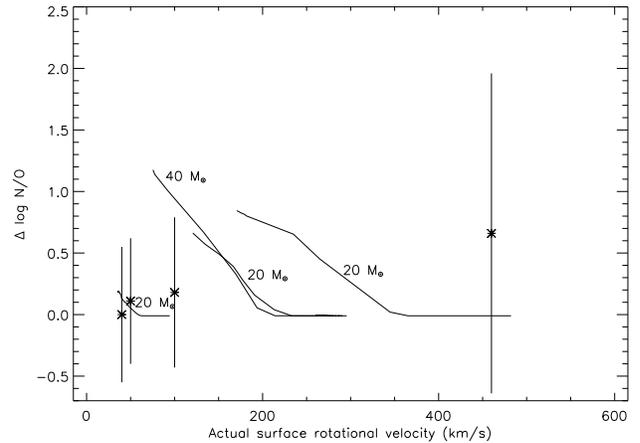}}
\caption[]{As fig. \ref{NCvrot} for the N/O ratio.}
\label{NOvrot}
\end{figure}

In order to obtain more quantitative conclusions we compare our N/C and
N/O ratios with the evolution of these quantities for certain evolutionary
models, in figs. \ref{NCvrot} and \ref{NOvrot} (Meynet 2001, 
private communication).

We refer the ratios to their initial values, that for the evolutionary
models is the solar composition and for our sample stars are those of 
the reference He normal stars.

Although we consider the CNO abundances of HD\,209975 as the reference 
unmixed values for low gravity stars, we show in figs. \ref{NCvrot} 
and \ref{NOvrot} its ratios refered to the reference dwarf HD\,214680. 

They are compatible with the evolution of a rotating star with initial
mass around 20 $M_{\sun}$, that has evolved until its actual (projected)
rotational velocity. Therefore, we leave open the question of the 
possible CNO contamination of this object.

For HD\,191423 we can only say that the star must have started its
main sequence lifetime with an extremely high rotational 
velocity, that leads to the present day projected value and surface
CNO contamination.

Therefore we find that it is possible to explain our results in the 
frame of massive stellar evolution with rotation, but we are still far 
from obtaining quantitative conclusions, due basically to the 
problem of the projection of the measurable rotational velocities
and to the small sample considered.

\section{Conclusions}\label{conclude}

We have studied a sample of four Galactic O9 stars in order to
investigate the relationship between stellar rotation and surface CNO
contamination.

The reference CNO abundances for unmixed O9 stars are given by the
three He normal stars studied, all located in the solar neighbourhood.
Their chemical compositions are indistinguishable within our
uncertainties. These abundances are in agreement with those
found in the solar neighbourhood for unmixed early B stars,
with values about 0.2 dex below solar.

The He and CNO abundances of the fast rotator HD\,191423 show the
trend expected from CNO contamination: He and N overabundance and C 
and O depletions. For the reference supergiant HD\,209975 we question
whether it shows also some surface contamination.

The comparison of our observed N/C and N/O ratios with the predictions
of the evolutionary models with rotation of Meynet \& Maeder
(\cite{Mey&Mae00}) show a qualitative agreement.

Nevertheless, we must be aware of the limitations of this result,
which represents only a first step in the investigation of the
relationship between stellar rotation and surface CNO contamination 
in massive O stars.

\begin{acknowledgements}
We wish to thank N. Przybilla, R.J. Garc\'\i a L\'opez, G. Meynet and 
N. Langer for providing us with data necessary for this work and for 
many valuable suggestions and discussions. AH acknowledges support for this
work by the spanish DGES under project PB97-1438-C02-01 and by the
Gobierno Aut\'onomo de Canarias under project PI1999/008.
\end{acknowledgements}



\end{document}